\documentclass[12pt]{iopart}
\usepackage{graphicx,amsfonts}
\pdfoutput=1
\expandafter\let\csname equation*\endcsname\relax
\expandafter\let\csname endequation*\endcsname\relax 
\usepackage{amsmath}
\usepackage{verbatim}

\usepackage{color}
\definecolor{darkblue}{rgb}{0,0,0.5}
\definecolor{darkred}{rgb}{0.5,0,0}
\usepackage[colorlinks=true,urlcolor=darkblue,citecolor=darkblue,linkcolor=darkred,hyperfootnotes=false]{hyperref}
\begin{document}

\newcommand{\atanh}
{\operatorname{atanh}}
\newcommand{\ArcTan}
{\operatorname{ArcTan}}
\newcommand{\ArcCoth}
{\operatorname{ArcCoth}}
\newcommand{\Erf}
{\operatorname{Erf}}
\newcommand{\Erfi}
{\operatorname{Erfi}}
\newcommand{\Ei}
{\operatorname{Ei}}

%commands for V
\newcommand{\eff}{{\text{eff}}}
\newcommand{\st}{\text{st}}
\newcommand{\eq}{\text{eq}}
\newcommand{\W}{\mathcal W}
\newcommand{\Ss}{\mathcal S}
\newcommand{\WW}{\mathbb W}
\newcommand{\T}{\mathcal T}
\newcommand{\Ww}{\mathcal W}
\newcommand{\Rr}{\mathcal R}

\newcommand{\qtot}{Q^{\tt tot}}
\newcommand{\qex}{Q^{\tt ex}}
\newcommand{\qhk}{Q^{\tt hk}}

\newcommand{\Stot}{S_{\tt tot}}

\newcommand{\fqtot}{\mathcal{Q}^{\tt tot}}
\newcommand{\fqex}{\mathcal{Q}^{\tt ex}}
\newcommand{\fqhk}{\mathcal{Q}^{\tt hk}}

\newcommand{\rhoss}{\rho_{\tt SS}}
\newcommand{\Yhasa}{\mathcal{Y}}

\newcommand{\C}{\mathcal C}
\newcommand{\R}{\text{R}}
\newcommand{\hk}{\text{hk}}
\newcommand{\tot}{\text{tot}}
\newcommand{\ad}{\text{a}}
\newcommand{\na}{\text{na}}
\newcommand{\ex}{\text{exp}}
\newcommand{\exc}{\text{exc}}
\newcommand{\StS}{\text{SS}}

\title{Joint probability distributions and fluctuation theorems}
\author{Reinaldo Garc{\'i}a-Garc{\'i}a$^1$\footnote{Corresponding author},
 Vivien Lecomte$^2$, Alejandro B. Kolton$^1$ and Daniel Dom{\'i}nguez$^1$}
\address{$^1$ Centro At\'omico Bariloche and Instituto Balseiro, 8400 S. C. de Bariloche, Argentina}
\address{$^2$ Laboratoire de Probabilit\'es et Mod\`eles Al\'eatoires (CNRS UMR 7599), Universit\'e Pierre et Marie Curie
 - Paris VI and Universit\'e Paris-Diderot - Paris VII,
 site Chevaleret, 175 rue du Chevaleret, 75013 Paris, France}
\ead{reinaldo.garcia@cab.cnea.gov.ar}

\begin{abstract}
We derive various exact results for 
Markovian systems that spontaneously relax to a
non-equilibrium steady-state by using joint probability distributions 
symmetries of different entropy production decompositions.
The analytical approach is applied to diverse problems such as 
the description of the fluctuations induced 
by experimental errors, for unveiling symmetries of correlation 
functions appearing in fluctuation-dissipation relations 
recently generalised to non-equilibrium steady-states, and also for 
mapping averages between different trajectory-based dynamical ensembles. 
Many known fluctuation theorems arise as special instances of our 
approach, for particular two-fold decompositions of the total entropy production.
  As a complement, we also briefly review and synthesise 
  the variety of fluctuation theorems applying to stochastic dynamics of both,
  continuous systems described by a Langevin dynamics and
  discrete systems obeying a Markov dynamics, emphasising how these results 
  emerge from distinct symmetries of the dynamical entropy of
  the trajectory followed by the system.
  For Langevin dynamics, we
  embed the ``dual dynamics'' with a physical meaning, and for Markov
  systems we show how the fluctuation theorems translate into
  symmetries of modified evolution operators.
\end{abstract}

\submitto{Journal of Statistical Mechanics}
\maketitle

\section{Introduction}
The study of the fluctuating heat interchange 
between a small system and a thermal reservoir is  
of academic interest but also of direct experimental relevance, as 
new techniques for microscopic manipulation and 
detection allow nowadays to measure fluctuations in small experimental 
systems of relevance in Physics, Chemistry and Biology~\cite{bustamante,ritort}. 
At this respect, a group of relations known as fluctuation theorems
(FT)~\cite{Evans-Cohen-Morris,Gallavotti-Cohen,Kurchan,Lebowitz,Jarzynski,Crooks,Hatano-Sasa,Seifert1,Seifert2} 
have attracted a lot of attention as they shed a new light into the principles governing energy 
fluctuations in a family of model systems. Remarkably, these 
results go beyond linear-response or quasi-equilibrium conditions, and apply to systems 
driven by non-conservative forces with arbitrary time-dependent protocols,  
even feedback controlled \cite{Horowitz,Sagawa,Seifert-feedback}.

Formally, the generality of the FTs can 
be attributed to the way probability distribution functions of particular observables 
behave under symmetry-breaking forcings, such as non-conservative and/or time 
dependent forces (see~\cite{Kurchan2,Harris-Schuetz} for reviews on FT).
Although the FTs are not expected to hold for every experimental 
system in contact with a thermal bath, 
they still provide 
a nice framework to analyse and understand the 
fluctuating heat interchange of out of equilibrium systems in general. 
Here we focus, as in most of the recent works in this field, on the family 
of systems whose driven dynamics can be well described 
by Langevin equations (without memory) or by discrete Markov chains. 
These are the prototypical models for which most FTs can be proven 
easily, without making any additional assumptions.
(Note that fluctuation theorems have also been derived 
for non-Markovian dynamics~\cite{Zamponi,Aron}.)

FTs are exact relations for the probability distributions for the
values $W$ of observables $\Ww[{\bf x};\sigma]$ which are functionals
of the stochastic state-space system trajectories ${\bf x}\equiv
\{x(t)\}_{t=0}^\tau$ (\emph{e.g.} work, heat or more generally,
different forms of trajectory-dependent entropy productions) in
processes driven by an arbitrary time-dependent protocol $\sigma(t)$
during a time $\tau$.  Typically, the so-called integral fluctuation
theorems (IFTs) are exact relations for the thermal average over
histories such as $\langle e^{- \Ww} \rangle = 1$ while detailed
fluctuations theorems (DFTs) for the same observable are stronger
relations, typically of the form $P(W)/P^T(-W)=e^W$, involving the
probability distributions $P(W)\equiv \langle \delta(\Ww[{\bf
  x};\sigma]-W)\rangle$ and $P^T(W)\equiv \langle \delta(\Ww[{\bf
  x};\sigma]-W)\rangle^T$ of the stochastic values $W$ of $\Ww$, with
$T$ denoting that the trajectories $x$ are sampled from a transformed
(typically time-reversed) dynamics such that $\Ww^T = -\Ww$.
% , 

The observable $\Ww$ is thus
 convenient to measure the asymmetry under the
transformation $T$. Indeed, $\langle W \rangle \equiv
D_{KL}(P(W) \parallel P^T(-W))$, with $D_{KL}(P_A \parallel P_B)$ the
Kullback-Leibler distance between two distributions $P_A$ and $P_B$.
Prominent examples for physically relevant observables $\Ww$ are the
total entropy production $\Stot$ which yields the Jarzynski IFT
\cite{Jarzynski} and Crooks DFT \cite{Crooks}, the non-adiabatic
entropy $S_{\na}$ produced in transitions between non-equilibrium
steady-states yielding the Hatano-Sasa IFT \cite{Hatano-Sasa} and DFT
\cite{Jarzynski1}, and the adiabatic entropy production $S_{\ad}$,
yielding the Speck-Seifert IFT \cite{Seifert1} and DFT
\cite{EspositoPRL}.

A simple unifying picture for all these seemingly different FTs 
has emerged recently \cite{Seifert2,EspositoPRL,Jarzynski1}. 
Although a DFT trivially includes as a particular case an IFT it is now clear 
that there exist two basic operations $T$ that we can use for generating the DFT version 
corresponding to each of the above mentioned IFTs. These two operations 
are {\it time-reversal} ($R$), which reverses the protocol maintaining the form of the dynamical 
equations, and the transformation to the 
so-called {\it dual dynamics} ($\dagger$), which corresponds to 
different dynamical equations.
These transformations are interesting since they are closely related to symmetry operations 
of equilibrium and non-equilibrium steady-states.
The above mentioned FTs can indeed be all unified in three detailed fluctuation 
theorems \cite{EspositoPRL,EspositoPRE1,EspositoPRE2} satisfied by the the total entropy production $\Stot$, 
and by each term of its particular two-fold decomposition, $\Stot = S_{\ad} + S_{\na}$, 
into an adiabatic $S_{\ad}$ and a non-adiabatic part $S_{\na}$. This splitting is 
physically motivated and closely related, for isothermal processes, 
to the splitting $Q_{\tot}=Q_{\hk}+Q_{\exc}$ of the total work done by all forces, 
into a house-keeping heat $Q_{\hk}$ and an excess heat $Q_{\exc}$, proposed by Oono 
and Paniconi \cite{Oono-Paniconi} and later formalised by Hatano-Sasa~\cite{Hatano-Sasa} 
for describing steady-state thermodynamics. 
The three DFT read
\begin{eqnarray}
 P(\Stot)/P^R(-\Stot) &=& e^{\Stot} \\
 P(S_{\na})/P^{\dagger R}(-S_{\na}) &=& e^{S_{\na}}, \\
 P(S_{\ad})/P^{\dagger}(-S_{\ad}) &=& e^{S_{\ad}} 
\label{eq:esposito3dft}
\end{eqnarray}
where in the second equation the dual and the the time-reversal operations 
are composed.  
As we can see, at variance with the DFT for $\Stot$, where the distribution of $\Stot$
for the forward process is compared with the one for the backward 
process obeying a time-reversed protocol, the detailed fluctuations 
theorems for $S_{\ad}$ and $S_{\na}$ require comparing the forward process with a 
process governed by the adjoint dynamics (which can 
be additionally time-reversed), which is very different from the 
original physical dynamics and therefore difficult to impose 
experimentally in general.

In a recent paper we have shown that writing detailed theorems 
in terms of joint probabilities is a convenient approach for deriving 
easily a family of 
relations, including the three detailed FTs, 
which might prove relevant for diverse applications. Introducing 
joint distributions is on the other hand natural when we are interested 
in the path dependent fluctuations of non-scalar observables. 
Indeed more general transformations $T$ such as vector rotations~\cite{Hurtado} 
can be also considered for deriving joint-distributions FTs.
In this work we provide more detailed calculations and expand the results 
of~\cite{Garcia-Garcia}, and then propose new applications.

The organisation of the paper is as follows. In Section~\ref{sec:rev}
we review various FTs and introduce the notations and basic
observables used in the next sections. In Section~\ref{sec:FTjoint} we
review the main results of our previous work, and give their detailed
derivation using Langevin dynamics in its path-integral
(Onsager-Machlup) representation (paragraph~\ref{ssec:langevin}) and
Markov chains (paragraph~\ref{ssec:markov}). In Section~\ref{sec:appli} we discuss
some applications of our results.  Conclusions and perspectives are
gathered in Section~\ref{sec:concl}.

\section{Fluctuation theorems preliminaries}
\label{sec:rev}

It is clear that given two different dynamical weights 
${\mathcal{P}}[{\bf x}; \sigma]$ and ${\mathcal{P}^T}[{\bf x}; \sigma]$ for the 
stochastic trajectories ${\bf x}\equiv \{x(t)\}_{t=0}^\tau$ of a given system, 
we can always define a trajectory dependent quantity
\begin{equation}
\Ww [{\bf x}; \sigma]=\ln(\mathcal{P}[{\bf x}; \sigma]/{\mathcal{P}^T}[{\bf x}; \sigma]). 
\end{equation}
which satisfies, by construction, the symmetry relation
\begin{equation}
\langle \mathcal{O}[{\bf x}; \sigma] \rangle_{p_0} = 
\langle \mathcal{O}[{\bf x}; \sigma] e^{\Ww[{\bf x};\sigma]}\rangle^{T}_{p_1}, 
\label{eq:dftgeneral}
\end{equation}
where $\langle...\rangle = \int \mathcal{D}{\bf x}{\mathcal{P}}[{\bf x}; \sigma]...$, 
$\langle...\rangle^T = \int \mathcal{D}{\bf x}{\mathcal{P}^T}[{\bf x}; \sigma]...$,  
with initial conditions sampled from the arbitrary distributions 
$p_0(x)$ and $p_1(x)$ respectively. The functional $\mathcal{O}[{\bf x},\sigma]$ 
is an arbitrary observable and thus equation (\ref{eq:dftgeneral}) can be used 
to get a detailed statistics. $\Ww$ is by construction odd under the swapping of the two dynamics, and 
its average over the first dynamics is the Kullback-Leiber distance between 
the two trajectory ensembles ${\mathcal{P}}[{\bf x}; \sigma]$ and ${\mathcal{P}^T}[{\bf x}; \sigma]$, that is 
$\langle \Ww \rangle = \mathcal{D}_{KL}({\mathcal{P}}[{\bf x}; \sigma]\parallel{\mathcal{P}^T}[{\bf x}; \sigma])$.
One may use the general equation (\ref{eq:dftgeneral}) in two interesting ways:
\begin{itemize}
 \item 
{\bf Mapping trajectory ensembles:} On one hand we might directly choose two different dynamics 
(through their transition probabilities or through their Langevin equations for instance) and use 
the equation to compare them quantitatively and from a purely information theoretical point of view. 
At this respect it is worth noting equation (\ref{eq:dftgeneral}) ``maps'' averages 
of arbitrary observables $\mathcal{O}$ in the original trajectory ensemble to another average in a 
``target ensemble'' of trajectories. An  interesting case is the one which maps averages 
in a true non-equilibrium ensemble where detailed balance is broken (e.g. driven 
by non-conservative forces) to a one satisfying detailed balance (e.g. driven by potential forces).  
A particular instance where $\Ww$ does acquires also a clear physical meaning 
is the mapping to equilibrium dynamics that is discussed in section~\ref{mappingtoeq}.
\item {\bf Symmetry transformations and entropy production:} On the other hand, 
instead of choosing a ``target'' dynamics,  we might focus on the properties of the original dynamics
and directly choose transformations $T$ of it connected 
to symmetries of their equilibrium and non-equilibrium steady-states (NESS). 
As it is well known and we briefly review in the next sections, what make this case 
specially interesting is that for such special transformations usually $\Ww$ acquires 
a well defined physical meaning as generalised trajectory entropy productions. 
As we show in the following sections, time-reversal and the transformation 
to the so called ``dual'' dynamics yield the three detailed theorems of
(\ref{eq:esposito3dft}) and thus many known fluctuations theorems of great 
relevance in stochastic thermodynamics.  
\end{itemize}

\subsection{Time-reversed dynamics}
\label{ssec:TR}
In the {\it time-reversed} dynamics, $T = R$, trajectories are still
governed by the original dynamical equations, with a time-reversed
protocol $\sigma^R(t) = \sigma(\tau-t)$.  When a system is driven out
of equilibrium, either by non-conservative or time dependent forces,
the resulting asymmetry is \emph{e.g.} measured by comparing the statistical
weights of two ``twin trajectories''~\cite{Jarzynski1} ${\bf x}$ and ${\bf x}^R$
such that their components are related as $x^R(t)=x(\tau-t)$, 
that is ${\bf x}$ is evolved with $\sigma(t)$ and and ${\bf x}^R$
with $\sigma^R(t)$. A convenient measure between the two weights is
the functional
\begin{equation}
\label{totaltrajentroprod}
\Ss [{\bf x}; \sigma]=\ln(\mathcal{P}[{\bf x}; \sigma]/\mathcal{P}^R[{\bf x}^R; \sigma^R]), 
\end{equation}
defined from the statistical weight $\mathcal{P}[{\bf x}; \sigma]$ of system
trajectories ${\bf x}$ evolved, starting from an initial condition
distribution {$\operatorname{Prob}[x(0)=s] = p_0(s)$}, during a given time interval
$\tau$ in a $d$-dimensional state space under the action of forces
controlled by an arbitrary set $\sigma \equiv \sigma(t)$ of
time-dependent external parameters~\cite{Kurchan,Lebowitz,Seifert2}.
Here $\mathcal{P}^R[{\bf x}^R; \sigma^R]$ denotes the
statistical weight of the trajectory ${\bf x}$ but evolved backwards,
$x^R(t)=x(t-\tau)$, and with the backward protocol
$\sigma^R(t)=\sigma(\tau-t)$, sampled from an initial condition distribution
$\operatorname{Prob}[x^R(0)=s] = p_1(s)$.  For instance, for a Langevin dynamics we
typically have $\mathcal{P}[{\bf x};\sigma]\sim e^{-\int_0^\tau
  dt\;\mathcal{L}[x,\dot x;\sigma]+\ln p_0(x(0))}$, and
$\mathcal{P}^R[{\bf x}^R;\sigma^R] \sim
e^{-\int_0^\tau dt\;\mathcal{L}[x^R,\dot x^R;\sigma^R]+\ln
  p_1(x^R(0))}$, where $\mathcal{L}[x,\dot x;\sigma]$ is the dynamical action
 of the system which, in general, also contains the logarithm of a Jacobian, 
 whose form depends on the considered stochastic calculus.
 By making a simple change of variables, we easily
derive $\Ss [{\bf x}; \sigma]\sim -\int_0^\tau dt\;(\mathcal{L}[x,\dot
x;\sigma]-\mathcal{L}[x,-\dot x;\sigma])-\ln (p_0(x(0))/p_1(x(\tau)))$
(see Section~\ref{ssec:langevin}). From \eqref{totaltrajentroprod} we have,
by construction, that $\Ss$ is odd under $T=R$, that is, $\Ss[{\bf x}^R; \sigma^R]=-\Ss[{\bf x}; \sigma]$, since $R$ is an
involution.  We also note that the definition of $\Ss$ allows us to
write, for the average of any observable $\mathcal{O}[{\bf x};\sigma]$
\begin{equation}
\label{eq:reversingtimeinaverage}
\langle \mathcal{O}[{\bf x}; \sigma]\rangle = \langle \mathcal{O}[{\bf x}^R; \sigma] e^{-S[{\bf x};\sigma^R]}  \rangle^R, 
\end{equation}
where $\langle...\rangle = \int \mathcal{D}{\bf x} \;\mathcal{P}[{\bf x};\sigma]...$ 
and $\langle...\rangle^{R} = \int \mathcal{D}{\bf x} \;\mathcal{P}^R[{\bf x};\sigma^R]...$ denotes the average over forward and 
reversed trajectories respectively. 
We note also that the concept of twin trajectories used to define $\Ss$ is irrelevant 
in the last equation since on each side we integrate over all possible trajectories (i.e. ${\bf x}$ is now a dummy variable) 
and that $\int\mathcal{D}{\bf x}^R=\int\mathcal{D}{\bf x}$ since $x^R$ is simply a time reflection and shift of the trajectory 
${\bf x}$ in time-state space.
Equation~\eqref{eq:reversingtimeinaverage} implies, in particular
\begin{equation}
%\label{dft_total}
\langle e^{-\lambda \Ss[{\bf x};\sigma] }\rangle=\langle e^{-(1-\lambda)\Ss[{\bf x};\sigma^R] }\rangle^R,
\end{equation}
where in the left hand side we recognise the generating function of
$P(S)=\langle \delta(\Ss[{\bf x};\sigma]-S) \rangle$
(we use calligraphic symbols to
differentiate functionals of stochastic trajectories from their actual
values) with $\lambda$ an arbitrary number we can use to compute any
cumulant of $S$. Introducing $P^R(S)=\langle \delta(\Ss[{\bf x};\sigma^R]-S) \rangle^R$,
 it is then straightforward to derive the detailed FT
(DFT),
\begin{equation}
\label{dft_total}
P(S)/P^R(-S)=e^S 
\end{equation}
which implies, by direct integration or by setting $\lambda=1$ above, the integral FT (IFT)
\begin{equation}
\label{ift_total}
\langle e^{-\Ss[{\bf x};\sigma] }\rangle = 1
\end{equation}
and thus, by using Jensen's inequality we get
\begin{equation}
\label{jensen_total}
\langle \Ss[{\bf x};\sigma] \rangle \geq 0.
\end{equation}
This inequality can also be obtained by noting that $\langle \Ss
\rangle$ is equal to the positively defined Kullback-Leibler distance
between two probability distributions, that is, for the present case,
$\langle \Ss \rangle= \int {\cal D}{\bf x}\; \mathcal{P}[{\bf x};\sigma]
\Ss[{\bf x};\sigma] = {\cal D}_{KL}(\mathcal{P}[{\bf x}; \sigma]||\mathcal{P}^R[{\bf x}^R;
\sigma^\R]) \ge 0$.  Since the equality above is thus reached for
time-reversal symmetric processes, at equilibrium $\langle \Ss \rangle=0$, as expected.  It is
easy to show that such symmetry is actually stronger, $\Ss=0$, due to
detailed balance.  In addition, if the relaxation time of the
equilibrium states is finite, then processes that are driven very
slowly compared with it do not produce this quantity either,  %% not ... either (and not neither)
as the
adiabatic process makes the system visit a sequence of the (symmetric)
equilibrium states compatible with the instantaneous value
$\sigma(t)$.  It is also worth remarking here that all the above
statistical properties are valid for arbitrary protocols
$\sigma(t)$, arbitrary initial conditions for the forward $p_0$ and
backward $p_1$ process, and they are time-independent, that is, they
are valid for any $\tau \ge 0$.  Let us note however that our
definition of $\Ss$ includes a border term containing
information about the initial condition , $p_0$ and $p_1$, of the twin
processes it compares. We note also that the choice of $p_1$ is free,
and not a constraint for the trajectories at the border $t=\tau$ 
however, a direct connection with ``physical''
entropy production $\Stot$ can be done if one chooses $p_1$ being the solution
of the Fokker-Planck (or master) equation at time $\tau$ 
\cite{Seifert2}.

What makes $\Ss$ physically interesting in connection with stochastic
thermodynamics is that it can be identified, up to a time border term,
with the stochastic heat dissipated into the reservoir divided by its
temperature along the stochastic system trajectory $x$, for an
isothermal process. If such process is described for instance by a
Langevin dynamics $\dot{x} = f(x;\sigma)+\xi$, where the force
$f(x;\sigma)$ might contain both conservative and non-conservative
terms, it can be shown that~\cite{Kurchan,SeifertEPJB} (see
paragraphs~\ref{ssec:langevin} and ~\ref{ssec:markov} for the
derivation in Langevin dynamics and Markov chains situations
respectively),
\begin{eqnarray}
\label{kurchan}
\Ss [{\bf x}; \sigma]&=& \frac{1}{T}\int_0^\tau dt\;f(x;\sigma) \dot{x} 
+ \ln \frac{p_0(x(0))}{p_1(x(\tau))} \\
&=& \frac{\qtot[{\bf x};\sigma]}{T} + \ln \frac{p_0(x(0))}{p_1(x(\tau))}
\end{eqnarray}
where the functional of the trajectory $\qtot[{\bf x};\sigma]$ is the total 
work done by the force $f(x;\sigma)$ 
on the stochastic trajectory $x$ under the protocol $\sigma$, 
and the border term $\ln [p_0(x(0))/p_1(x(\tau)]$ 
depends only on the time boundaries of each stochastic 
trajectory, sampled by $p_0$ and $p_1$ which remain
so far arbitrary.
With this identification of $\Ss$, it is also interesting to note again 
that since $\langle \Ss \rangle = {\cal D}_{KL}(\mathcal{P}[{\bf x}; \sigma]||\mathcal{P}^R[{\bf x}^R; \sigma^\R])$, 
~(\ref{jensen_total}) and~(\ref{kurchan}) relate irreversibility and 
dissipation in an elegant information theoretical way. At this respect it 
is worth noting that in equilibrium $\sigma$ is constant, 
$p_{0,1}(x)=e^{-\beta (E(x;\sigma)-F(\sigma))}$, and since all forces are 
conservative, we have $\qtot = E[x(0),\sigma]-E[x(\tau),\sigma]$, yielding 
$\Ss [{\bf x}; \sigma]=0$ for each stochastic trajectory 
${\bf x}$. By writing $\Ss$ in terms of transition probabilities one obtains that 
this is equivalent to  the detailed balance condition.

Equations (\ref{eq:reversingtimeinaverage}-\ref{kurchan}) can be combined to derive many known FTs for Markovian systems. 
To start, by combining  (\ref{eq:reversingtimeinaverage}) and (\ref{kurchan}), one
obtains the generalised Crooks relation for the average of an arbitrary 
observable $\mathcal{O}[{\bf x};\sigma]$ along a trajectory with the forward and 
backward protocol,
\begin{equation}
\left\langle \mathcal{O}[{\bf x};\sigma] 
e^{-\frac{1}{T}{\qtot[{\bf x};\sigma]} - \ln \frac{p_0(x(0);\sigma(0))}{p_1(x(\tau);\sigma(\tau))}} 
\right\rangle_{p_0} = 
\left\langle \mathcal{O}[{\bf x};\sigma] 
\right\rangle^R_{p_1}.  
\end{equation}
We have now a lot of freedom to derive FTs, by choosing appropriate values for $\mathcal{O}$, $p_0$ and $p_1$. 
\begin{itemize}
\item {\it Seifert relation}: Choosing $\mathcal{O}=1$ and
  $p_1(x)=\rho(x,\tau)$ to be the time-dependent solution of the
  Fokker-Planck equation with initial condition $p_0(x)\equiv
  \rho(x,0)$, we obtain the Seifert theorem~\cite{Seifert2}, valid for all times
  $\tau$ and arbitrary initial conditions. This choice defines the
  so-called {\it trajectory dependent total entropy production} $\Stot
  = \qtot/T + S_s$, with $S_s = \ln [\rho(x,\tau)/\rho(x,0)]$ the {\it
    trajectory dependent system entropy production}, such that:
\begin{equation}
\label{ift_total_siefert}
\left\langle e^{-\Stot} \right\rangle_{\text{any} \;p_0} = 1.
\end{equation}
Seifert also made the interesting observation that this choice for $p_1$ is optimal in the sense 
$\min_{p_1} [\langle \Ss \rangle_{p_0}]=\langle \Ss_{\tot} \rangle_{p_0}$. 
This is easy to understand 
if we write
\begin{eqnarray}
\langle \Ss - \Ss_{\tot} \rangle_{p_0}=
\left\langle \ln \frac{\rho(x(\tau),\tau)}{p_1(x(\tau))} \right\rangle_{p_0} =\nonumber\\
=\int dy\; \rho(y,\tau) \ln \frac{\rho(y,\tau)}{p_1(y)} =
\mathcal{D}_{KL}( \rho(y,\tau) \parallel p_1(y)) 
\ge 0,
\end{eqnarray}
and use the positiveness of the Kullback-Leiber distance between any two distributions.
\item {\it Jarzynski relation for NESS}: 
By choosing $\mathcal{O}=1$, $p_1(x)=\rho_{\StS}(x;\sigma(\tau))$ and $p_0(x)=\rho_{\StS}(x;\sigma(0))$
from the steady-state solution of the Fokker-Planck equation $\rho_{\StS}(x;\sigma)$
we obtain the generalised Jarzynski relation for transitions between 
NESS, valid for all times $\tau$ and steady-state initial conditions compatible 
with $\sigma(0)$. 
\begin{equation}
\label{ift_jarzynski_gral}
\left\langle e^{-\frac{\qtot[{\bf x};\sigma]}{T} - [\phi(x(\tau),\sigma(\tau))-\phi(x(0);\sigma(0))]} 
\right\rangle_{\rhoss(\sigma(0))} = 1.
\end{equation}
where $\phi(x;\sigma) \equiv -\ln \rhoss(x;\sigma)$. As noted by Hatano and Sasa 
this relation does not generalise the second law of thermodynamics for NESS, since its corresponding Jensen's inequality
\begin{equation}
\left\langle \frac{\qtot[{\bf x};\sigma]}{T} + [\phi(x(\tau),\sigma(\tau))-\phi(x(0);\sigma(0))]
\right\rangle_{\rhoss(\sigma(0))} \geq 0,
\end{equation}
does not reach zero for adiabatic processes due to the presence of non-conservative forces 
which, even in the steady-state, inject energy and produce the so-called house-keeping heat. 
\item {\it Jarzynski relation}: 
The IFT of equation \eqref{ift_jarzynski_gral}
reduces to the well known Jarzynski relation 
if the initial steady-state 
is the Boltzmann-Gibbs equilibrium state $\rho_{\StS}(x;\sigma) = \rho_{\eq}(x;\sigma) = e^{-\beta [E(x;\sigma)-F(\sigma)]}$, with 
$E(x;\sigma)$ the energy of state $x$ and $F(\sigma)$ the free energy, under the constraint $\sigma$
\begin{equation}
\label{ift_jarzynski}
\left\langle e^{-\beta{\Ww_d}} \right\rangle_{\rho_{\eq}(\sigma(0))} = e^{-\beta{\Delta F} }
\end{equation}
where the dissipated work is $\Ww_d[{\bf x};\sigma] \equiv \qtot[x;\sigma] + E[x(\tau);x(\tau)]-E[x(0);\sigma(0)]$ 
and $\Delta F = F(\sigma(\tau))-F(\sigma(0))$. Again:
\begin{equation}
\left\langle {\Ww_d} \right\rangle_{\rho_{\eq}(\sigma(0))} \ge \Delta F.
\end{equation}
The equality is achieved in the adiabatic limit, and thus Jensen's inequality yields
 the second law for transitions between equilibrium states. Note that (an infinite number) of other
second law-like inequalities can be obtained from variational methods~\cite{infinite2ndlawlike}.
\item {\it Crooks relation}:
If $p_0(x)=\rho_{\eq}(x;\sigma(0))$ and $p_1(x)=\rho_{\eq}(x;\sigma(\tau))$, 
and $\mathcal{O}[{\bf x};\sigma]= \delta(\Ww_d[{\bf x};\sigma]-W_d) $,  
equation (\ref{eq:reversingtimeinaverage}) reduces to
\begin{equation}
P(W_d) e^{-\beta \Delta F}= \nonumber \\ 
P^R(-W_d) e^{-W_d}
\end{equation}
where $P(W_d)=\langle \delta(\Ww_d[{\bf x};\sigma]-W_d) \rangle$, 
$P^R(W_d)=\langle \delta(\Ww_d^R[{\bf x};\sigma]-W_d) \rangle^R$,
\item {\it Fluctuation in NESS}: In the absence of time-dependent forces, 
$\sigma(t)=\sigma_0$, a system initially prepared in the steady-state 
$p_0(x) = \rhoss(x;\sigma_0)$ remains there and satisfy 
a particular form of the Crooks relation:
\begin{equation}
P(W_d) = P(-W_d) e^{-W_d}
\end{equation}
\end{itemize}

\subsection{Dual dynamics}
Non-equilibrium 
steady-states (NESS) are already asymmetric with respect to time-reversal due to the lack of 
detailed balance. This has led Oono and Paniconi~\cite{Oono-Paniconi} to introduce a useful two fold 
decomposition of the total heat exchange into a ``house-keeping 
heat'' part, constantly produced to maintain the non-equilibrium 
driven steady-state with non-vanishing currents, and an ``excess heat'' part, 
produced only during transitions between steady-states. 
The excess heat is minimised in adiabatic processes, that is when $\sigma(t)$ varies 
very slowly compared with an assumed finite relaxation time towards the NESS, 
while the house keeping heat is minimised only at equilibrium, in the absence of 
non-conservative forces, when detailed balance is recovered. 
Hatano and Sasa formalised this splitting for Langevin dynamics by defining stochastic 
trajectory dependent quantities, $\qhk[{\bf x};\sigma]$ and $\qex[{\bf x};\sigma]$, 
for the house-keeping and excess heat respectively, and deriving 
a IFT which generalised the second law for transitions between NESS. 
Although they do not use it for its derivation, they point out that 
the so-called dual dynamics, denoted by the symbol $\dagger$, 
composed with time-reversal, plays a role analogous to time-reversal 
alone in the derivation of the Jarzynski equality.

Indeed, the adjoint transformation is defined such that NESS are symmetric with respect to the simultaneous 
application of time-reversal $R$ and the adjoint transformation $\dagger$ to 
the original dynamics, that is, with respect to the composed ``$T=R \circ \dagger$'' transformation: 
the steady-state of the adjoint dynamics has the same distribution $\rhoss^\dagger=\rhoss$ 
as the original dynamics but with an inverted steady-state current $J_{\StS}^\dagger = - J_{\StS}$. 
Since the original current can be recovered by a time-reversal without changing $\rhoss$, 
the steady-state is symmetric with respect to the composed operation $T=R \circ \dagger$.
It is worth
remarking that the adjoint transformation changes the dynamics (see
paragraph~\ref{ssec:dualdynCT} for an explicit example). 
It is then natural to introduce the so-called  trajectory-dependent non-adiabatic 
entropy production
\begin{equation}
\label{eq:defSna}
\Ss_{\na} [{\bf x}; \sigma]=\ln(\mathcal{P}[{\bf x}; \sigma]/\mathcal{P}^{\dagger R}[{\bf x}^R; \sigma^R]), 
\end{equation}
to measure the asymmetry produced when the system is driven out of the
NESS.  
Here $\mathcal{P}^{\dagger R}[{\bf x}^R; \sigma^R]$ is the weight of the
trajectory ${\bf x}$ in the time-reversed dual dynamics. In Section~\ref{sec:FTjoint} we derive explicit
forms for $S_{\na}$ for Langevin and Markov chain dynamics. For
Langevin dynamics we get
\begin{eqnarray}
\label{ssna}
\Ss_{\na} [{\bf x}; \sigma]&=& -\int_0^\tau dt\;\frac{\partial\phi}{\partial x}(x;\sigma) \dot{x} 
+ \ln \frac{p_0(x(0))}{p_1(x(\tau))} \\
&=& \frac{\qex[{\bf x};\sigma]}{T} + \ln \frac{p_0(x(0))}{p_1(x(\tau))}
\end{eqnarray}
making a physical connection with the Oono-Paniconi-Hatano-Sasa excess heat. Note that here 
we have used again the freedom to choose the initial condition distributions $p_0$ and $p_1$ for 
the ``twin trajectories'', the first weighted in the forward protocol 
of the physical dynamics, the second weighted in the backward protocol of the dual dynamics. 
By analogy, it is now straightforward to 
write FTs. The definition of $\Ss_{\na}$ allows us to write, for the average of any observable 
$\mathcal{O}[{\bf x};\sigma]$, the FT generator

\begin{equation}
\langle \mathcal{O}[{\bf x}; \sigma] e^{-\Ss_{\na}[{\bf x};\sigma]} \rangle_{p_0} 
= \langle \mathcal{O}[{\bf x}; \sigma] \rangle^{R\dagger}_{p_1}, 
\end{equation}
where $\langle...\rangle^{\dagger R} = \int \mathcal{D}{\bf x} \;\mathcal{P}^{\dagger R}[{\bf x};\sigma^R]...$ 
denotes the average over trajectories generated by the dual dynamics and controlled by a time-reversed protocol 
and $\Ss_{\na}^\dagger[{\bf x}^{R};\sigma^R]=-\Ss_{\na}[{\bf x};\sigma]$.
It is worth remarking here that the above statistical relation is valid for any 
initial condition distributions $p_0$ and $p_1$ and is  valid for any $\tau \ge 0$.
We can now easily derive several relations by making particular choices for $\mathcal{O}$, and $p_{0,1}$:
\begin{itemize}
 \item {\it Generating function}:
By choosing $\mathcal{O}[{\bf x};\sigma] = e^{(1-\lambda) \Ss_{\na}[{\bf x};\sigma]}$, with $\lambda$ any number, we get
\begin{equation}
\label{dft_nonadiabatic}
\langle e^{-\lambda \Ss_{\na}[{\bf x};\sigma] }\rangle_{p_0}
=\langle e^{-(1-\lambda)\Ss_{\na}[{\bf x};\sigma^R] }\rangle^{R \dagger}_{p_1},
\end{equation}
\item {\it Non-adiabatic entropy production DFT}:  By choosing 
$\mathcal{O}[{\bf x};\sigma] = \delta(\Ss_{\na}[{\bf x};\sigma]-S)$ and defining
$P(S)=\langle \delta(\Ss_{\na}[{\bf x};\sigma]-S) \rangle_{p_0}$ and  
$P^{R\dagger}(S)=
\langle \delta(\Ss_{\na}^{\dagger R}[{\bf x};\sigma]-S) \rangle^{R\dagger}_{p_1}
= \langle \delta(\Ss_{\na}[{\bf x};\sigma]+S) \rangle^{R\dagger}_{p_1}
$, 
it is straightforward to derive the DFT~\cite{Jarzynski1,EspositoPRL}
\begin{equation}
P(S_{\na})/P^{\R\dagger}(-S_{\na})=e^{S_{\na}} 
\end{equation}
which is the detailed version of the Hatano-Sasa IFT, if the initial condition is the stationary distribution.
\item {\it Hatano-Sasa IFT}: 
By choosing $\mathcal{O}[{\bf x};\sigma]=1$ and 
$p_0(x)=\rhoss(x;\sigma(0))$ and $p_1(x)=\rhoss(x;\sigma(\tau))$ we obtain 
the Hatano-Sasa IFT
\begin{equation}
\label{ift_nonadiabatic}
\langle e^{-(\beta \qex[{\bf x};\sigma]+\Delta \phi )}\rangle_{\rhoss(\sigma(0))} = 1
\end{equation}
where $\Delta \phi = \phi(x(\tau);\sigma(\tau))-\phi(x(0);\sigma(0))$. By using Jensen's inequality we get
\begin{equation}
\label{jensen_nonadiabatic}
\beta \langle \qex[{\bf x};\sigma] \rangle_{\rhoss(\sigma(0))} \geq \langle \Delta \phi \rangle_{\rhoss(\sigma(0))}.
\end{equation}
which is the generalisation of the second-law for transition between
NESS.  Again, this inequality can also be obtained by noting that
$\langle \Ss_{\na} \rangle$ is equal to the positively defined
Kullback-Leibler distance between two probability distributions.  For
the present case, $\langle \Ss \rangle= {\cal D}_{KL}(\mathcal{P}[{\bf x};
\sigma]||\mathcal{P}^{\dagger R}[{\bf x}^R; \sigma^R]) \ge 0$, and the
equality is reached for processes that are time-reversal symmetric in
the dual dynamics.  Therefore equilibrium in particular, and NESS in
general, have $\langle \Ss \rangle=0$, as expected.  The latter is
also true for adiabatic processes, slow compared to an assumed finite
relaxation time toward the NESS, so that the system is always very
close to the NESS corresponding to the instantaneous value of
$\sigma(t)$. The absence of non-adiabatic entropy production in NESS
is actually ``detailed'': $\Ss_{\na}=0$. As shown in
Section~\ref{ssec:dualdynCT} this can be understood from the
definition of dual dynamics and from a detailed balance like relation
between the transition probabilities of the direct and dual dynamics
for the same pair of states.
\end{itemize}

Finally, we note that equilibrium states are symmetric with respect to $R$ and $\dagger$. 
It is thus natural to introduce a new quantity $\Ss_{\ad}$ to measure the asymmetry 
produced by non-conservative or time-dependent driving forces by using the $\dagger$ operation alone, 
\begin{equation}
\label{eq:defSa}
\Ss_{\ad} [{\bf x}; \sigma]=\ln(\mathcal{P}[{\bf x}; \sigma]/\mathcal{P}^\dagger[{\bf x}; \sigma]). 
\end{equation}
where we note that the twin trajectories are actually the same, the first weighted in 
the direct dynamics and the second in the dual dynamics.
In Section~\ref{sec:FTjoint} we derive an explicit form for $S_{\ad}$, from Langevin and Markov 
chain dynamics. For Langevin dynamics the following relation holds %, for $p_0(x(0))\neq p_1(x(0))$ (it holds also for p0=p1, no?? Vivien)
\begin{eqnarray}
\Ss_{\ad} [{\bf x}; \sigma]&=& \frac{1}{T}\int_0^\tau dt\;v_{\StS}(x;\sigma)  \dot{x} 
+ \ln \frac{p_0(x(0))}{p_1(x(0))} \\
&=& \frac{\qhk[{\bf x};\sigma]}{T}+\ln \frac{p_0(x(0))}{p_1(x(0))}
\end{eqnarray}
where the steady-state velocity $v_{\StS}(x;\sigma) \equiv J_{\StS}(\sigma)/\rho_{\StS}(x;\sigma)$, 
with $\rho_{\StS}(x;\sigma)\equiv e^{-\phi(x;\sigma)}$ the NESS probability distribution, and $J_{\StS}$ the probability current 
in state-space. We also note that the initial condition for trajectories weighted in the dual 
dynamics $p_1$ is not reversed.
The above makes the physical connection with the house-keeping heat. 
By analogy with the previous cases, it is now straightforward to 
write a FT generator for this observable: 
\begin{equation}
\langle \mathcal{O}[{\bf x}; \sigma] e^{-\Ss_{\ad}[{\bf x};\sigma]} \rangle_{p_0} = \langle \mathcal{O}[{\bf x}; \sigma] \rangle^{\dagger}_{p_1}, 
\end{equation}
where $\langle...\rangle^{\dagger} = \int \mathcal{D}x \;\mathcal{P}^{\dagger}[{\bf x};\sigma]...$ 
denotes the average over trajectories weighted in the dual dynamics. We can thus proceed analogously.
\begin{itemize}
 \item {\it Generating function}:
\begin{equation}
\langle e^{-\lambda \Ss_{\ad}[{\bf x};\sigma] }\rangle_{p_0}=\langle e^{-(1-\lambda)\Ss^\dagger_{\ad}[{\bf x};\sigma] }\rangle^{\dagger}_{p_1},
\end{equation}
\item {\it Adiabatic entropy production DFT}:
\begin{equation}
P(S_{\ad})/P^{\dagger}(-S_{\ad})=e^{S_{\ad}} 
\end{equation}
which is a detailed version of the Speck-Seifert IFT, if the initial condition is stationary.
\item {\it Speck-Seifert IFT}:
By choosing $\mathcal{O}[{\bf x};\sigma]=1$ and 
$p_0(x)=p_1(x) = \rhoss(x;\sigma(0))$ we obtain 
the Speck-Seifert IFT
\begin{equation}
\langle e^{-\beta \qhk[{\bf x};\sigma] }\rangle_{\rhoss(x;\sigma(0))} = 1
\end{equation}
by using Jensen's inequality we get
\begin{equation}
\label{jensen_adiabatic}
\langle \beta \qhk[{\bf x};\sigma] \rangle_{\rhoss(x;\sigma(0))} \geq 0.
\end{equation}
Again, this inequality can also be obtained by noting 
that $\langle \Ss_{\ad} \rangle$ is equal to the positively defined 
Kullback-Leibler distance between two probability distributions, 
that is, for the present case, 
$\langle \Ss_{\ad} \rangle= {\cal D}_{KL}(\mathcal{P}[{\bf x}; \sigma]||\mathcal{P}^{\dagger}[{\bf x}; \sigma]) \ge 0$.
Since the equality above is reached for dual-symmetric processes by construction, 
only equilibrium states have $\langle \qhk \rangle=0$. NESS do produce $\qhk$ 
because they need house-keeping energy to maintain detailed balance violation. In other words, NESS are 
dual asymmetric because the $\dagger$ operation, although keeping $\rhoss$ invariant, 
invert the steady-state current $J_{\StS}$. 
Therefore, the equality in equation \eqref{jensen_adiabatic} is never reached by NESS. 
At equilibrium, the equality is actually reached, and is ``detailed'' in the sense that
$\qhk[{\bf x};\sigma]=0$ for each trajectory, since
$J_{\StS}=0$ exactly. 

\end{itemize}

\subsection{Splitting}
By using the operations $R$ and $\dagger$, related to symmetries 
of equilibrium and NESS states, we have defined the trajectory dependent 
total, non-adiabatic, and adiabatic entropy production functionals, 
\begin{eqnarray}
\mathcal{P}[{\bf x};\sigma] &=& \mathcal{P}^R[{\bf x}^R;\sigma^R] e^{\Ss[{\bf x};\sigma]} \\
\mathcal{P}[{\bf x};\sigma] &=& \mathcal{P}^{\dagger R}[{\bf x}^R;\sigma^R] e^{\Ss_{\na}[{\bf x};\sigma]} \\
\mathcal{P}[{\bf x};\sigma] &=& \mathcal{P}^\dagger[{\bf x};\sigma] e^{\Ss_{\ad}[{\bf x};\sigma]}. 
\end{eqnarray}
From the last equation we have
\begin{eqnarray}
\mathcal{P}^{\dagger R}[{\bf x}^R;\sigma^R] &=& \mathcal{P}^R[{\bf x}^R;\sigma^R] e^{\Ss_{\ad}^{\dagger}[{\bf x}^R;\sigma^R]} 
\end{eqnarray}
therefore, from the second we get
\begin{eqnarray}
\mathcal{P}[{\bf x};\sigma] 
= \mathcal{P}^R[{\bf x}^R;\sigma^R] e^{\Ss_{\ad}^{\dagger}[{\bf x}^R;\sigma^R]+\Ss_{\na}[{\bf x};\sigma]} . 
\end{eqnarray}
Comparing with the first we conclude:
\begin{equation}
 \Ss[{\bf x};\sigma] = \Ss_{\ad}^{\dagger}[{\bf x}^R;\sigma^R] + \Ss_{\na}[{\bf x};\sigma]
\end{equation}
only using transformation properties. 
If we define $\Ss_{\ad}$ such that it does not include 
a border term as it is customary,
we see that $\Ss_{\ad}^{\dagger}[{\bf x}^R;\sigma^R]=\Ss_{\ad}[{\bf x};\sigma]$. 
This is so because $R$ and $\dagger$ only 
change the sign of $\Ss_{\ad}$. Therefore, we get 
\begin{equation}
 \Ss = \Ss_{\ad} + \Ss_{\na}
\end{equation}
which is the starting point of the three detailed theorems \cite{EspositoPRL}. However, it  is worth noting
that in \cite{EspositoPRL}, $\mathcal{S}_{\tt tot}$ have been considered instead of $\Ss$.
This splitting  is explicitly obtained for Langevin 
and Markov chains in section~\ref{sec:FTjoint}.

\subsection{Fluctuation theorems from joint distribution symmetries}
\label{sec:FTjoint}

In a previous work~\cite{Garcia-Garcia} we have shown that the use of
joint probability distributions for different total entropy production 
decompositions is a convenient tool for deriving a variety of exact 
expressions for Markovian systems, including many known 
fluctuations theorems arising from particular two-fold decompositions 
of the total entropy production.
This was done by first noting that {\it
  any} decomposition of the total trajectory entropy production for
Markov systems, $\Ss [{\bf x}; \sigma]=\sum_{i=1}^M \mathcal A_i[{\bf x}; \sigma]$,
has a joint probability distribution satisfying a generalised detailed
fluctuation theorem, when all the contributing terms are odd with
respect to time reversal, 
$\mathcal A_i[{\bf x}^\R; \sigma^\R]=-\mathcal A_i[{\bf x}; \sigma]$:
\begin{equation}
\frac{P(A_1, A_2,..., A_M)}{P^\R(-A_1, -A_2,..., -A_M)}=e^{S}
\label{eq:joint0}
\end{equation}
with ${S}=\sum_{i=1}^M A_i$ and $P(A_1, A_2,..., A_M)=\langle \delta(\mathcal{A}_1-A_1)...\delta(\mathcal{A}_M-A_M) \rangle_{p_0}$.
This contains the same information as
\begin{equation}
\left\langle e^{-\sum_{i=1}^M \lambda_i \mathcal A_i[{\bf x};\sigma]}\right\rangle_{p_0} 
= \left\langle e^{-\sum_{i=1}^M (1-\lambda_i) \mathcal A_i[{\bf x};\sigma^R]} \right\rangle^R_{p_1},
\end{equation}
for $\{\lambda_i\}$ $M$ arbitrary numbers. It is worth remarking that relations of the kind of the previous equation,
have been already derived in \cite{Lebowitz, Andrieux-Gaspard} for time independent $\sigma$.

More generally, for a transformation $T$, we can define a quantity 
${\Ss_T}=\ln \{\mathcal{P}[{\bf x};\sigma]/\mathcal{P}^T[{\bf x};\sigma]\}$.
If we can write $\Ss_T [{\bf x}; \sigma]=\sum_{i=1}^M \mathcal{B}_i[{\bf x}; \sigma]$ 
such that each component is odd or even under $T$, 
$\mathcal{B}^T_i = \epsilon^T_i \mathcal{B}_i$ with $\epsilon^T_i = \pm 1$, we get
\begin{equation}
\frac{P(B_1, B_2,..., B_M)}{P^T(\epsilon^T_1 B_1, \epsilon^T_2 B_2,..., \epsilon^T_M B_M)}=e^{\Ss_T}
\end{equation}
or, equivalently
\begin{equation}
\left\langle e^{-\sum_{i=1}^M \lambda_i \mathcal B_i[{\bf x};\sigma]}\right\rangle_{p_0} 
= \left\langle e^{\sum_{i=1}^M (1-\lambda_i) \epsilon^T_i \mathcal B_i^T[{\bf x};\sigma]} \right\rangle^T_{p_1}.
\end{equation}
On the other hand, for a not constrained list of variables $\{\mathcal{C}_i\}_{i=2}^M$, such that 
$\mathcal{C}^T_i = \epsilon^T_i \mathcal{C}_i$, we can also write
\begin{equation}
\left\langle e^{-\lambda \Ss_T +\sum_{i=2}^M \lambda_i \mathcal C_i[{\bf x};\sigma]}\right\rangle_{p_0} 
= \left\langle e^{-(1-\lambda)\Ss_T^T + \sum_{i=2}^M \lambda_i \epsilon^T_i \mathcal C_i^T[{\bf x};\sigma]} \right\rangle^T_{p_1},
\end{equation}
or, equivalently
\begin{equation}
\frac{P(S_T, C_2, C_3,..., C_M)}{P^T(-S_T, \epsilon^T_2 C_2, \epsilon^T_3 C_3,..., \epsilon^T_M C_M)}=e^{S_T}
\label{eq:generaljointDFT}
\end{equation}
These relations are interesting as they contain the three detailed fluctuation theorems \cite{EspositoPRL} 
as a particular case for $M=1$: For the three transformations $T= (R),\; (\dagger),\; (R \circ \dagger)$ we get, respectively, 
the DFTs for $\Ss$, $\Ss_a$ and $\Ss_{na}$. 
As an example, using the above relations for $M=2$ we can get the following useful identities for $\Ss_a$ and $\Ss_{na}$:
\begin{equation}
\fl{P(\Ss_a, \Ss_{na} )}={P^\dagger(-\Ss_a, \Ss_{na})} e^{\Ss_a}
={P^{\dagger R}(\Ss_a, -\Ss_{na})} e^{\Ss_{na}}
={P^{R}(-\Ss_a, -\Ss_{na})} e^{\Ss}
\end{equation}
The last one assuming the splitting $\Ss=\Ss_a+\Ss_{na}$.

Another interesting consequence 
of (\ref{eq:generaljointDFT}) and Bayes theorem is that  
\begin{equation}
{P(C_2,..., C_M|S_T)}={P^T(C_2^T,..., C_M^T|-S_T)}
\label{eq:generalconditionaljointDFT}
\end{equation}
meaning that the variables $\{\mathcal{C}_i\}_{i=2}^M$ have identical statistical 
properties in the trajectory subensembles determined by the constraints $\Ss_T[{\bf x},\sigma] = S_T$ 
and $\Ss_T^T[{\bf x},\sigma] = -S_T$, 
and can thus not be used to differentiate the original and the transformed dynamics. 
It is particularly instructive to consider variables $\{\mathcal{C}_i\}_{i=2}^M$ describing a discretized path of 
a certain duration. Following the discussion for the time-reversal case in Ref.\cite{gomezmarin} we can now 
write, for the particular transformations $T= (R),\; (\dagger),\; (R \circ \dagger)$, the subensembles equivalences 
\begin{eqnarray}
{P(\text{path}|S)}={P^R(\text{path}^R|-S)}\\
{P(\text{path}|S_a)}={P^\dagger(\text{path}|-S_a)}\\
{P(\text{path}|S_{na})}={P^{\dagger R}(\text{path}^R|-S_{na})}
\end{eqnarray}
As discussed in Ref.\cite{gomezmarin}, and can be appreciated recalling (\ref{eq:esposito3dft}),
the above expressions ``map'' the statistics of typical trajectories in 
one forward process, to rare trajectories in the transformed process, being the latter  backward, 
the backward-dual or the forward-dual process, according to the corresponding involved transformation.

\section{Markovian dynamics models}
So far we have made mostly general mathematical considerations arising 
from the definitions of trajectory entropy productions without specifying the dynamics
behind the corresponding and different trajectory statistical weights. In this section 
we analyse two paradigmatic class of Markov dynamics. First we consider 
the Langevin dynamics for systems with configurations lying in the continuum, 
and described by first order stochastic differential equations. Second we consider
the continuous-time Markov chains describing dynamical systems with discrete configurations,
described by master equations with well defined transition probabilities. 
\subsection{Langevin dynamics}
\label{ssec:langevin}
\subsubsection{Generalities}
For simplicity we consider in the following discussion the case of 
one dimensional Langevin system coupled with a single thermal bath. Our model consists in a 
Brownian particle driven by an external force. In the presence of periodic boundary conditions, 
the steady-state for this system
has a non-zero probability current. Our results can be easily generalised to more dimensions and many particles
(provided the temperature is the same along every spatial dimension~-- see \emph{e.g.}~\cite{Puglisi} for a physically
relevant situation where it is not the case).
We also consider in~\ref{ssec:markov} the case of discrete Markov chains. We start with the Langevin equation
\begin{equation}
\label{langevindynamics}
\dot{x} =-\frac{\partial U}{\partial x}(x;\alpha)+f+\xi, 
\end{equation}
where $\alpha$ represents a set of parameters of the system, $f$ is a driving force and $\xi(t)$ is a Gaussian white noise with
variance $\langle\xi(t)\xi(t')\rangle=2T\delta(t-t')$. The parameters $\alpha$ and $f$ may depend on time
and $\sigma=(\alpha,f)$. For this system
the probability of a given trajectory in the phase space has the following form
\begin{equation}
 \label{weight}
\mathcal{P}[{\bf x};\sigma]=\int\mathcal{D}\xi
P[\xi]p_0(x(0);\sigma_{0})J[{\bf x};\sigma]\delta\bigg[\dot{x}+\frac{\partial U}{\partial x}(x;\alpha)-f-\xi\bigg],
\end{equation}
where $p_0$ is the initial probability density function of the system (at $t=0$), $P[\xi]$ is the probability
distribution of the thermal noise and $J[{\bf x};\sigma]$ is a Jacobian to be defined below. $P[\xi]$ takes the following form
\begin{equation}
\label{noise}
 P[\xi]=\Pi^{-1}(\tau)\exp\bigg[-\frac{1}{4T}\int_{0}^{\tau}dt\,\xi(t)^{2}\bigg],
\end{equation} 
where $\Pi(\tau)$ is a normalisation constant.

It is instructive to make at this point a formal, but important
remark. It is essential to specify the discretization scheme
(\textit{i.e.} the choice of It\=o or Stratonovich stochastic
calculus~--- see~\cite{LauLubensky} for a detailed analysis) of the path
integral given in equation (\ref{weight}) and all the following
ones. In general, the relevant quantities we are interested in
(different forms of heat) were originally defined in the Stratonovich
scheme, and we work using this picture. Besides, the Stratonovich
discretization is easier to use in our context since it is invariant
under time-reversal, contrary to the It\=o one.
In the Stratonovich case, the Jacobian in
equation (\ref{weight}) reads
\begin{equation}
\label{jacobian}
J[{\bf x};\sigma]=\exp\bigg\{\frac{1}{2}\int_{0}^{\tau}dt\frac{\partial^{2}U}{\partial x^{2}}(x;\alpha)\bigg\}. 
\end{equation}
As regards time-reversal transformations, an important property of
this Jacobian is the following:
\begin{equation}\label{jacobiansymmetry}
J[{\bf x};\sigma]=J[{\bf x}_R;\sigma_{R}].
\end{equation}
 Integrating out the noise in equation (\ref{weight}) and taking into account equation (\ref{noise}) we obtain:
\begin{equation}
\label{weight1}
\mathcal{P}[{\bf x},\sigma]=\Pi^{-1}(\tau)J[{\bf x};\sigma]\exp\{-I[{\bf x};\sigma]\},
\end{equation}
where the Onsager-Machlup action functional $I[{\bf x};\sigma]$  writes
\begin{equation}
 \label{action}
I[{\bf x};\sigma] =-\ln p_0(x(0);\sigma_0)+\int_{0}^{\tau}dt\bigg\{\frac{1}{4T}
\bigg[\dot{x}^{2}+\bigg(\frac{\partial U}{\partial x}(x;\alpha)-f\bigg)^{2}\bigg]
+\frac{1}{2T}\dot{x}\bigg(\frac{\partial U}{\partial x}(x;\alpha)-f\bigg)\bigg\}.
\end{equation}
Let us now consider the time reversed probability weight given by
\begin{equation}
\label{weightreversed}
\mathcal{P}^R[{\bf x}_R,\sigma_R]=\Pi^{-1}(\tau)J[{\bf x}_R;\sigma_R]\exp\{-I_R[{\bf x}_R;\sigma_R]\}, 
\end{equation}
with
\begin{eqnarray}
\label{actionR}
\fl I_R[{\bf x}_R;\sigma_R]=-\ln p_1(x(\tau);\sigma_\tau)+\int_{0}^{\tau}dt\bigg\{\frac{1}{4T}
\bigg[(\dot{x}_R)^{2}+\bigg(\frac{\partial U}{\partial x}(x_R;\alpha_R)-f_R\bigg)^{2}\bigg]+\nonumber\\
\quad\qquad\qquad\qquad +\frac{1}{2T}\dot{x}_R\bigg(\frac{\partial U}{\partial x}(x_R;\alpha_R)-f_R\bigg)\bigg\},
\end{eqnarray}
where $p_1$ is from now on the solution of the Fokker-Planck equation
of the process at time $\tau$ in order to make a connection with~\cite{Seifert2}. In this case one can immediately write
for the total trajectory entropy production
\begin{equation}
 \label{entropyprodlangevin}
\mathcal{S}[{\bf x}; \sigma]=\ln
(\mathcal{P}[{\bf x}; \sigma]/\mathcal{P}^R[{\bf x}_R; \sigma_R])=I_R[{\bf x}_R;\sigma_R]-I[{\bf x};\sigma]\equiv
\mathcal{S}_s[{\bf x};\sigma]+\mathcal{S}_r[{\bf x};\sigma],
\end{equation}
where one identifies the system (reservoir) entropy production $\mathcal{S}_s[{\bf x};\sigma]$
($\mathcal{S}_r[{\bf x};\sigma]$) as
\begin{equation}
 \label{systementropy}
\mathcal{S}_s[{\bf x};\sigma]=-\ln\frac{ p_1(x(\tau);\sigma_\tau)}{ p_0(x(0);\sigma_0)},
\end{equation}
\begin{equation}
 \label{reservoirentropy}
\mathcal{S}_r[{\bf x};\sigma]=-\frac{1}{T}\int_{0}^{\tau}dt\dot{x}\bigg(\frac{\partial U}{\partial x}(x;\alpha)-f\bigg).
\end{equation}
Consider now the instantaneous stationary probability density function
for given values of the set of parameters of the system
$\rho_{\StS}(x(t);\sigma(t))=e^{-\phi(x(t);\sigma(t))}$ and let us add
and subtract the quantity $\int_0^\tau
dt\dot{x}\frac{\partial\phi}{\partial x}(x;\sigma)$ from
equation (\ref{entropyprodlangevin}). In this case one finds a different
decomposition of the total trajectory entropy production in two
different contributions, the so-called adiabatic and the non-adiabatic
contributions
\begin{equation}
 \label{adiabaticentropy}
\mathcal{S}_{\ad}=\int_{0}^{\tau}dt\dot{x}\bigg[\frac{\partial\phi}{\partial x}(x;\sigma)
-\frac{1}{T}\bigg(\frac{\partial U}{\partial x}(x;\alpha)-f\bigg)\bigg],
\end{equation}
\begin{equation}
 \label{nonadiabaticentropy}
\mathcal{S}_{\na}=\ln\frac{ p_0(x(0);\sigma_0)}{ p_1(x(\tau);\sigma_\tau)}
-\int_0^\tau dt\dot{x}\frac{\partial\phi}{\partial x}(x;\sigma).
\end{equation}
We thus have two relevant decompositions
\begin{equation}
 \label{entropy1}
\mathcal{S}[{\bf x}; \sigma]=\mathcal{S}_s[{\bf x};\sigma]+\mathcal{S}_r[{\bf x};\sigma]=\mathcal{S}_{\ad}[{\bf x};\sigma]+\mathcal{S}_{\na}[{\bf x};\sigma].
\end{equation}
Note that instead of using the Onsager-Machlup action functional $I[{\bf x};\sigma]$
one could equivalently work in the Martin-Siggia-Rose-Janssen-De Dominicis framework
(which is more suitable for non-Markovian dissipation~\cite{Aron}).

\subsubsection{Dual dynamics in continuous time}
\label{ssec:dualdynCT}
When the steady-state probability density function for a
given system doesn't satisfies detailed balance, one usually
introduces the dual dynamics in terms of its propagator. In a time
discretized picture one defines it via the relation:
\begin{equation}
\label{dualdef}
K^{\dagger}(x_{i}|x_{i+1};\sigma_{i})=K(x_{i+1}|x_{i};\sigma_{i})\frac{\rho_{\StS}(x_{i};\sigma_{i})}
{\rho_{\StS}(x_{i+1};\sigma_{i})}. 
\end{equation}
As can be seen, when detailed balance holds
the dual propagator is equal to the corresponding propagator for the dynamics of the
system.

In the limit of continuous time we have that $x_{i}=x(s), \; \sigma_{i}=\sigma(s), 
\; x_{i+1}=x(s+ds), \; ds\rightarrow 0$. In that case
\begin{equation}
\label{propagatorcontinuous}
K(x(s+ds)|x(s);\sigma(s))\simeq\exp\bigg[\frac{1}{2}ds\frac{\partial^{2}U}{\partial x^{2}}(x;\alpha)-
\frac{1}{4T}ds\bigg(\dot{x}+\frac{\partial U}{\partial x}(x;\alpha)-f\bigg)^{2}\bigg]. 
\end{equation}
From (\ref{dualdef}) and (\ref{propagatorcontinuous}) we have:
\begin{eqnarray}
\label{dualcontinuous}
\fl K^{\dagger}(x(s+ds)|x(s);\sigma(s))
\simeq\exp\bigg[\frac{1}{2}ds\frac{\partial^{2}U}{\partial x^{2}}(x;\alpha)-
\frac{1}{4T}ds\bigg(\dot{x}-\frac{\partial U}{\partial x}(x;\alpha)+f\bigg)^{2}-\nonumber\\
\qquad\qquad\qquad -ds\dot{x}\frac{\partial\phi}{\partial x}(x;\sigma)\bigg]. 
\end{eqnarray}
Now, from (\ref{dualcontinuous}) we obtain for the general dual propagator:
\begin{equation}
\label{dualcontinuous1}
K^{\dagger}(x,t|x',t';\sigma)=\Pi^{-1}(t-t')\int_{x(t')=x'}^{x(t)=x}DxJ[{\bf x};\sigma]\exp[-\mathcal{I}^{\dagger}[{\bf x};\sigma]],
\end{equation}
where
\begin{equation}
\label{dualaction}
\mathcal{I}^{\dagger}[{\bf x};\sigma]=\int_{t'}^{t}ds\bigg\{\frac{\beta}{4}\bigg[\dot{x}(s)-\frac{\partial U}{\partial x}
(x(s);\alpha(s))+f(s)\bigg]^{2}+
\dot{x}\frac{\partial\phi}{\partial x}(x(s);\sigma(s))\bigg\}
\end{equation}
and $J[{\bf x};\sigma]$ is given by equation (\ref{jacobian}).  Before going
beyond, let us write the action (\ref{dualaction}) in a more
clear way.  By adding and subtracting the quantity
$\frac{1}{T}\dot{x}(\frac{\partial U}{\partial x}-f)$ in the integrand
of equation (\ref{dualaction}), we obtain:
\begin{equation}
\label{dualaction1}
\mathcal{I}^{\dagger}[{\bf x};\sigma]=\int_{t'}^{t}ds\bigg\{\frac{1}{4T}\bigg[\dot{x}(s)+\frac{\partial U}{\partial x}(x(s);\alpha(s))-f(s)\bigg]
^{2}\bigg\}+
\mathcal{S}_{\ad}[{\bf x};\sigma].
\end{equation}
We thus confirm the identification of $\Ss_{\ad}$ with the trajectory relative entropy between its weight in the 
original dynamics with its weight in the dual dynamics, as defined in~\eqref{eq:defSa}:  
\begin{equation}
\label{eq:linkSaPPdagger}
\mathcal{S}_{\ad}[{\bf x};\sigma] = \mathcal{I}^{\dagger}[{\bf x};\sigma]-\mathcal{I}[{\bf x};\sigma] = 
\ln \frac{P[{\bf x};\sigma]}{P^{\dagger}[{\bf x};\sigma]}
\end{equation}
The second equality in the previous equation holds for $p_0(x(0))=p_1(x(0))$.
 From this remark one obtains as an immediate result the
validity of an IFT for the adiabatic entropy production
\begin{equation}
 \label{IFTSa}
\langle e^{-\mathcal{S}_{\ad}[{\bf x};\sigma]}\rangle=\langle 1\rangle^\dagger\equiv1.
\end{equation}
To endow the dual dynamics with a physical meaning, one would
like to associate the dual propagator to an effective Langevin
equation (\textit{i.e.} an effective microscopic dynamics).  To do so,
let us consider the following action:
\begin{equation}
\label{effectiveaction}
\mathcal{I}_{\eff}[{\bf x},\sigma]= \int_{t'}^{t}ds\frac{1}{4T}\bigg[\dot{x}+\frac{2}{\beta}
\frac{\partial\phi}{\partial x}(x;\sigma)-\frac{\partial U}{\partial x}(x;\alpha)+f\bigg]^{2}.
\end{equation}
Simple algebra shows that:
\begin{equation}
\label{dual-effective relation}
 \mathcal{I}_{\eff}[{\bf x},\sigma]=
\mathcal{I}^{\dagger}[{\bf x};\sigma]-\int_{t'}^{t}ds\bigg[\frac{\partial\phi}{\partial x}(x;\sigma)
\bigg(\frac{\partial U}{\partial x}(x;\alpha)-f\bigg)-T\bigg(\frac{\partial\phi}{\partial x}(x;\sigma)\bigg)^{2}\bigg].
\end{equation}
We know that, by definition, $\rho_{\StS}(x,\sigma)$ satisfies at fixed
$\sigma$ the following stationary Fokker-Planck equation:
\begin{equation}
\label{fokker-planck}
T\frac{\partial^{2}\rho_{\StS}}{\partial x^{2}}(x;\sigma)+\frac{\partial}{\partial x}
\bigg[\bigg(\frac{\partial U}{\partial x}(x;\alpha)-f\bigg)\rho_{\StS}(x;\sigma)\bigg]=0.
\end{equation}
Now, putting $\rho_{\StS}(x;\sigma)=\exp[-\phi(x;\sigma)]$ in (\ref{fokker-planck}) one obtains:
\begin{equation}
\bigg(\frac{\partial\phi}{\partial x}\bigg)\bigg(\frac{\partial U}{\partial x}-f\bigg)
-T\bigg(\frac{\partial\phi}{\partial x}\bigg)^{2}=
\frac{\partial^{2}}{\partial x^{2}}(U-T\phi). 
\end{equation}
We thus get:
\begin{equation}\label{action-jacobian}
 \mathcal{I}_{\eff}[{\bf x};\sigma]=\mathcal{I}^{\dagger}[{\bf x};\sigma]+\frac{1}{2}\int_{t'}^{t}ds\frac{\partial^{2}}{\partial x^{2}}
\bigg[2T\phi(x;\sigma)-2U(x;\alpha)\bigg].
\end{equation}
Substituting (\ref{action-jacobian}) in (\ref{dualcontinuous1})
yields:
\begin{equation}\label{dualcontinuous2}
K^{\dagger}(x,t|x',t')=\Pi^{-1}(t-t')\int_{x(t')=x'}^{x(t)=x}DxJ_{\eff}[{\bf x};\sigma]\exp[-\mathcal{I}_{\eff}[{\bf x};\sigma]],
\end{equation}
where:
\begin{eqnarray}
\label{effectivejacobian}
J_{\eff}[{\bf x};\sigma] &=& J[{\bf x};\sigma]\exp\bigg\{\frac{1}{2}\int_{t'}^{t}ds\frac{\partial^{2}}{\partial x^{2}}
\bigg[\frac{2}{\beta}\phi(x;\sigma)-2U(x;\alpha)\bigg]\bigg\}=\nonumber\\
 &=& \exp\bigg\{\frac{1}{2}\int_{t'}^{t}ds
\frac{\partial^{2}}{\partial x^{2}}\bigg[\frac{2}{\beta}\phi(x;\sigma)-U(x;\alpha)\bigg]\bigg\}.
\end{eqnarray}
Let us define now the effective potential:
\begin{equation}\label{effectivepotential}
V_{\eff}(x;\sigma)=\frac{2}{\beta}\phi(x;\sigma)-U(x;\alpha). 
\end{equation}
With this definition  we have that:
\begin{equation}\label{effectiverelations}
\mathcal{I}_{\eff}[{\bf x};\sigma]=\bigg[\dot{x}+\frac{\partial}{\partial x}V_{\eff}(x;\sigma)+f\bigg]^{2},\quad %\nonumber \\
J_{\eff}[{\bf x};\sigma]=\exp\bigg[\frac{1}{2}\int_{t'}^{t}ds\frac{\partial^{2}}{\partial x^{2}}V_{\eff}(x;\sigma)\bigg]. 
\end{equation}
From (\ref{dualcontinuous2}) and (\ref{effectiverelations}) one obtains 
that the dual propagator is the propagator corresponding, in the Stratonovich scheme, to the Langevin
equation:
\begin{equation}\label{duallangevin}
\dot{x}=-\frac{\partial}{\partial x}V_{\eff}(x;\sigma)-f+\xi, 
\end{equation}
where, as can be seen from (\ref{langevindynamics}), the force has reversed sign. It is easy now to see that when
 detailed balance holds, this dynamics coincides with the 
real dynamics (\ref{langevindynamics}). In fact, in this case we have
 $\phi(x;\sigma)=\frac{1}{T}[U(x)-fx-F(T;\sigma)]$, where $F(T;\sigma)$ is the free energy.
Substitution of this choice for $\phi$ in (\ref{duallangevin}) directly leads to (\ref{langevindynamics}).
Concluding this discussion, we remark that the dual dynamics
corresponds to the dynamics of a system having the same steady-state
probability density function but with \emph{opposite} probability
current in the steady-state.

Finally we can also compute $\Ss_{\na}$ directly from the path-integral representation of the Langevin dynamics as   
\begin{eqnarray}
\Ss_{\na} = I_R^{\dagger}[{\bf x}^R;\sigma^R]-I[{\bf x};\sigma] &=& \ln \frac{p_{0}(x(t');\sigma(t'))}{p_1(x(t);\sigma(t))}+
\mathcal{I}^{\dagger}[{\bf x}^R;\sigma^R]-\mathcal{I}[{\bf x};\sigma]=\nonumber\\
 &=& \ln \frac{p_{0}(x(t');\sigma(t'))}{p_1(x(t);\sigma(t))}-\int_{t'}^t ds\;\frac{\partial\phi}{\partial x}(x;\sigma) \dot{x}.
\end{eqnarray}
Taking now $t'=0$ and $t=\tau$, we obtain (\ref{ssna}).

\subsubsection{Derivation of known FTs}
We now use the general symmetry equation~\eqref{eq:joint0} and the two
relevant decompositions obtained above for the trajectory entropy
production in the considered case in order to reobtain, from an
unified point of view, the known FTs. First one sees immediately that
\begin{equation}
 \label{jointprobs}
P(S_s,S_{r})/P^R(-S_s,-S_{r})=e^{S_s+S_{r}} \quad \mbox{and} \quad
P(S_{\ad},S_{\na})/P^R(-S_{\ad},-S_{\na})=e^{S_{\ad}+S_{\na}},
\end{equation}
or equivalently
\begin{equation}
 \label{jointprobs1}
\langle e^{-\lambda_1\mathcal{S}_s-\lambda_2\mathcal{S}_r}\rangle=
\langle e^{-(1-\lambda_1)\mathcal{S}^R_s-(1-\lambda_2)\mathcal{S}^R_r}\rangle_R 
\quad \mbox{and} \quad
\langle e^{-\lambda_1\mathcal{S}_{\ad}-\lambda_2\mathcal{S}_{\na}}\rangle=
\langle e^{-(1-\lambda_1)\mathcal{S}^R_{\ad}-(1-\lambda_2)\mathcal{S}^R_{\na}}\rangle_R.
\end{equation}
It is worth noting
that these relations do not involve dual probability distribution functions (PDFs), and thus they can be
tested for a physical system with a given dynamics. We also note that
while one can show both that $S_{\ad}$ and $S_{\na}$ satisfy separately
a DFT by using dual PDFs~\cite{EspositoPRL}, $S_s$ and $S_r$ satisfy a
joint DFT although they do not satisfy separately a DFT.

Let us now derive from a unified view the known FTs. Using equation (\ref{eq:linkSaPPdagger}) we can introduce
the dual joint probability density function
\begin{equation}
\label{dualjoint}
 P^\dagger(S_{\ad}, S_{\na})=P(-S_{\ad}, S_{\na})e^{S_{\ad}}.
\end{equation}
 This probability density function is, by virtue of equation (\ref{IFTSa}), correctly normalised
\begin{eqnarray}
 \label{norm}
\int dS_{\ad} dS_{\na}P^\dagger(S_{\ad}, S_{\na})=\int dS_{\ad} dS_{\na}P(-S_{\ad}, S_{\na})e^{S_{\ad}}=\nonumber\\
=\int dS_{\ad} dS_{\na}P(S_{\ad}, S_{\na})e^{-S_{\ad}}=\langle e^{-S_{\ad}}\rangle=1.
\end{eqnarray}
Let us first derive the Speck-Seifert DFT \cite{Seifert1}. We have
\begin{equation}
 \label{speck-seifert}
P(S_{\ad})=\int dS_{\na}P(S_{\ad}, S_{\na})=\int dS_{\na}P^\dagger(-S_{\ad}, S_{\na})e^{S_{\ad}}=P^\dagger(-S_{\ad})e^{S_{\ad}}
\end{equation}
We can also derive the Chernyak-Chertkov-Jarzynski DFT \cite{Jarzynski1}
for the non-adiabatic contribution
\begin{eqnarray}
 \label{CCJ}
P(S_{\na})=\int dS_{\ad}P(S_{\ad}, S_{\na})=\int dS_{\ad}P^R(-S_{\ad},-S_{\na})e^{S_{\ad}+S_{\na}}=\nonumber\\
=\int dS_{\ad}P^{\dagger R}(S_{\ad},-S_{\na})e^{S_{\na}}=P^{\dagger R}(-S_{\na})e^{S_{\na}}.
\end{eqnarray}

\subsection{Markov dynamics}
\label{ssec:markov}

\subsubsection{Settings}

\begin{figure}[ht]
  \centering
  \includegraphics{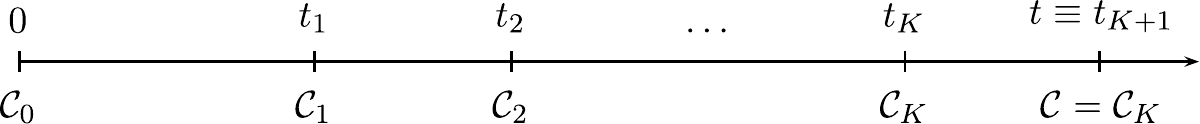}
  \caption{A history of configurations $\C_0\to\ldots\to\C_K$. Between
    $t_{k}$ and $t_{k+1}$, the system remains in configuration
    $\C_k$. }
  \label{fig:time_sequence}
\end{figure}

The symmetry~\eqref{eq:joint0} holds for any decomposition of the
entropy as a sum of terms which are antisymmetric by time
reversal. The difficult step is to explicitly write such kinds of
decompositions. In this section, we consider a continuous time Markov
process for a system described by discrete configurations $\{\C\}$,
and work out such decompositions for that dynamics. The master
equation for the probability $P(\C,t)$ for the system to be in
configuration $\C$ at time $t$ writes
\begin{equation} \label{eqn:cont_time_Markov_eq}
  \partial_t P(\C,t) = 
    \sum_{\C'}\big[ W(\C'\to\C,\sigma)P(\C',t)
                                   -  W(\C\to\C',\sigma)P(\C,t)  \big]
\end{equation}
where $\sigma$ is an external time-dependent control parameter.  A
system history consists of a sequence $\C_0,\ldots,\C_K$ of
configurations with jumps at times $t_1,\ldots,t_K$ (see
Fig \ref{fig:time_sequence}).

Let us introduce the dual transition rates $W^\dag$
\begin{equation}
  W^\dag(\C\to\C',\sigma) 
 \equiv \frac{P_\st(\C',\sigma)}{P_\st(\C,\sigma)} W(\C'\to\C,\sigma)
 = e^{-[\phi(\C',\sigma)-\phi(\C,\sigma)]} W(\C'\to\C,\sigma) 
\end{equation}
Note that when detailed balance is obeyed (\emph{i.e.}
$P_\eq(\C)W(\C\to\C')=P_\eq(\C')W(\C'\to\C)$ for some equilibrium
distribution $P_\eq(\C)$), the dual dynamics is the same as the original one.
We first remark that the dual escape rate 
$
 r^\dag(\C,\sigma)\equiv \sum_{\C'} W^\dag(\C\to\C',\sigma)
$
is the same as the original one:
\begin{equation}
  r^\dag(\C,\sigma)=\frac 1{P_\st(\C,\sigma)} \sum_{\C'}
  {W(\C'\to\C,\sigma) P_\st(\C',\sigma)}
  = \frac 1{P_\st(\C,\sigma)} r(\C,\sigma)P_\st(\C,\sigma)
  =r(\C,\sigma)
\end{equation}
More importantly, the \emph{steady-state} of the dual dynamics is the same since
\begin{eqnarray}
  \sum_{\C'}W^\dag(\C'\to\C,\sigma)P_\st(\C',\sigma)=
  \sum_{\C'}       W(\C\to\C',\sigma)P_\st(\C,\sigma) =\nonumber\\
= r(\C,\sigma) P_\st(\C,\sigma) = r^\dag(\C,\sigma)  P_\st(\C,\sigma)
\end{eqnarray}

\subsubsection{Entropies}

The (density of) probability of a trajectory specified as in Figure~\ref{fig:time_sequence} writes
\begin{equation}
 \operatorname{Prob}[\C,\sigma] =  e^{-\int_0^\tau dt\, r(\C(t),\sigma(t))}
  \prod_{k=1}^K W(\C_{k-1}\to\C_k,\sigma_{t_k})\ P_0(\C(0),\sigma(0))
\end{equation}
where $K$ is the number of events and $P_{0}$ the initial
distribution.  It precisely means that the average of an observable
$\mathcal O$ depending on the history of configurations and on the
protocol $\sigma$ writes
\begin{equation}
  \langle\mathcal O\rangle
= \sum_{K\geq 0}
 \sum_{\C_{0}\ldots \C_{K}}
 \int_0^t dt_K \int_0^{t_K} dt_{K-1} \ldots \int_0^{t_2} dt_1 \:
 \mathcal O[\C,\sigma]
\:
 \operatorname{Prob}[\C,\sigma]
\end{equation}
By analogy to systems described by Langevin dynamics, one defines the total entropy
\begin{align}
 \Ss[\C,\sigma] 
 &= \log \frac{ \operatorname{Prob}[\C,\sigma] }{ \operatorname{Prob}^R[\C^\R,\sigma^\R] }
\\
 &= \sum_{k=1}^K \log \frac{ W(\C_{k-1}\to\C_k,\sigma_{t_k})}{W(\C_{k}\to\C_{k-1},\sigma_{t_k})} 
 \ + \ \log \frac{ P_{0}(\C(0),\sigma(0))}{ P_{1}(\C(t),\sigma(t))}
\label{eq:def_total-entropy}
\end{align}
where $\C^\R$ is the reversed trajectory and $\sigma^\R$ the reverse
protocol.  Note that $\Ss[\C,\sigma] =-\Ss[\C^\R,\sigma^\R]$.  We now
would like to split the action into a sum of different terms, each of
them also \emph{antisymmetric} upon time-reversal.  To do so, we
assume that $P_{0}=P_{1}$ is the steady
state $P_{\text{st}}=e^{-\phi}$. Then, we define the `house-keeping'
entropy $\mathcal Q^\hk$ as
\begin{align}
 \beta \mathcal Q^\hk[\C,\sigma] 
&=
 \sum_{k=1}^K \log\frac{W(\C_{k-1}\to\C_k,\sigma_{t_k})}{W^\dag(\C_{k-1}\to\C_k,\sigma_{t_k})}
\\
&= 
 \sum_{k=1}^K \log\frac{W(\C_{k-1}\to\C_k,\sigma_{t_k})}{W(\C_k\to\C_{k-1},\sigma_{t_k})}
 + 
 \sum_{k=1}^K\phi(\C_k,\sigma_{t_{k}}) - \phi(\C_{k-1},\sigma_{t_k})
\end{align}
note that it is indeed antisymmetric upon time reversal: $ \beta
\mathcal Q^\hk[\C,\sigma] =- \beta \mathcal Q^\hk[\C^\R,\sigma^\R]$.
Moreover, we see by direct computation (\emph{e.g.}
from~\eqref{eq:HSfunctional_decomposed} below), that the total entropy
$\Ss$ writes in terms of the house-keeping work as
\begin{equation}
%\boxed
{
 \Ss[\C,\sigma]
=
 \beta \mathcal Q^\hk[\C,\sigma]
+
 \mathcal Y [\C,\sigma]
}
\end{equation}
where 
\begin{equation}
  \mathcal Y [\C,\sigma] = \int_0^\tau dt\:\dot\sigma  \frac{\partial\phi}{\partial\sigma}
\end{equation}  
is the Hatano-Sasa functional. Note that each term of the decomposition is antisymmetric.

Note last that defining $\Delta\phi=\phi(\C(t),\sigma(t))-\phi(\C(0),\sigma(0))$ and
\begin{equation}
 \beta \mathcal Q^\tot[\C,\sigma]
 = 
 \sum_{k=1}^K \log \frac{ W(\C_{k-1}\to\C_k,\sigma_{t_k})}{W(\C_{k}\to\C_{k-1},\sigma_{t_k})} 
\end{equation}
one reads directly from the definition~\eqref{eq:def_total-entropy} that
\begin{equation}
%\boxed
{
 \Ss[\C,\sigma]
=
 \beta \mathcal Q^\tot[\C,\sigma]
+
 \Delta\phi
}
\end{equation}
where again each term of this decomposition is antisymmetric.

To summarise:
\begin{eqnarray}
 \Ss[\C,\sigma]
=
\overbrace{\vphantom{\sum_{k=1}^K} \beta \mathcal Q^\tot[\C,\sigma]}^{\Ss_1}
\ + \ 
\overbrace{\sum_{k=1}^K\big[\phi(\C_k,\sigma_{t_{k}}) - \phi(\C_{k-1},\sigma_{t_k})\big]}^{\Ss_2}
\ - \ \nonumber\\
\qquad\qquad-\overbrace{\sum_{k=1}^K\big[\phi(\C_k,\sigma_{t_{k}}) - \phi(\C_{k-1},\sigma_{t_k})\big]}^{\Ss_2}
\ + \ 
\overbrace{\vphantom{\sum_{k=1}^K}\Delta\phi}^{\Ss_3}
\label{eq:decompositionSMarkov}
\end{eqnarray}
The first decomposition consists in grouping $\beta \mathcal Q^\hk[\C,\sigma]=\Ss_1+\Ss_2$
and $\mathcal Y[\C,\sigma]=-\Ss_2+\Ss_3$
while the second decompositions simply corresponds to  $\beta \mathcal Q^\tot[\C,\sigma]=\Ss_1$
and $\Delta\phi=\Ss_3$.

We also remark that writing
\begin{equation}
 \Ss[\C,\sigma]
= \beta \mathcal Q^\hk[\C,\sigma]
\ - \ 
\sum_{k=1}^K\big[\phi(\C_k,\sigma_{t_{k}}) - \phi(\C_{k-1},\sigma_{t_k})\big]
\ + \ 
\Delta\phi
\end{equation}
one has a decomposition in a sum of \emph{three} antisymmetric terms implying FTs using~\eqref{eq:joint0}

\subsubsection{Link to the Hatano-Sasa functional and symmetries of operators}

It is instructive to rewrite the Hatano-Sasa functional $\mathcal Y$ as
\begin{align}
 \mathcal Y[\C,\sigma]&=\int_0^\tau dt\:\dot\sigma  \frac{\partial\phi}{\partial\sigma}
=\sum_{k=0}^K\int_{t_k}^{t_{k+1}} dt\:\dot\sigma  \frac{\partial\phi}{\partial\sigma}
 =\sum_{k=0}^K\phi(\C_k,\sigma_{t_{k+1}}) - \phi(\C_k,\sigma_{t_k})
\nonumber \\
 &= \big[\phi(\C,\sigma)\big]_0^\tau - \sum_{k=1}^K\phi(\C_k,\sigma_{t_{k}}) - \phi(\C_{k-1},\sigma_{t_k})
\label{eq:HSfunctional_decomposed}
\end{align}
We thus have split $\mathcal Y$ in two parts,
$\big[\phi(\C,\sigma)\big]_0^\tau$, which depends explicitly on the
final time $\tau$, and the reduced Hatano-Sasa functional
\begin{align}
 \hat h_\tau = - \sum_{k=1}^K\phi(\C_k,\sigma_{t_{k}}) - \phi(\C_{k-1},\sigma_{t_k})  
 \label{eq:def_hat_h}
\end{align}
which (for fixed protocol $\sigma(t)$) depends only on the sequence of
visited configurations $\C_k$ and of the jump times $t_k$.  This
decomposition helps us to write $\langle e^{-s  \mathcal Y}\rangle$ in
terms of an $s$-modified evolution operator.
We show in appendix~\ref{app:symm_operators} that the DFTs arising
from the the decomposition~\eqref{eq:decompositionSMarkov} of the
dynamical entropy translate into symmetries
of a modified operators of evolutions $\mathbb W_\lambda$ such that
$\langle e^{-\lambda S}\rangle=\langle P_1|\mathcal T e^{t \mathbb W_\lambda}|P_0\rangle$
where $\mathcal T e^\cdot$ is the time-ordered exponential.
Compared to the original approach of Hatano and
Sasa~\cite{Hatano-Sasa} our derivations bring only into play the
properties of the operator of evolution (through the use of the
time-ordered exponential) and do not require a time discretization.

\section{Applications}
\label{sec:appli}
\subsection{Experimental errors}
\subsubsection{General derivation}
We now show with some detail how joint FTs provide insights on the
experimental error in the evaluation of entropy productions. The
results we have shown up to now are rather general and do not depend
on the shape of the initial and final PDFs for the considered systems
under arbitrary protocols. For concreteness and in order to be closer
to experiments, we consider in this section the case of transitions
between NESS, so the initial and final PDFs for the system, are taken
as the ones corresponding to the steady-states of the system with the
given values of the parameters.

In this situation, the non-adiabatic contribution to the total trajectory entropy production is simply the 
Hatano-Sasa (H-S) functional~\cite{Hatano-Sasa}: 
\begin{equation}
 \label{hasa}
\mathcal{Y}[{\bf x};\sigma]=\int_0^\tau dt\dot{\sigma}\frac{\partial\phi}{\partial\sigma}(x;\sigma)
\end{equation}
We are interested in experimental measurements of this quantity,
performed for instance in~\cite{Trepagnier}. In a typical experiment
one follows in general the following protocol:
 
\begin{itemize}
\item First one measures the particle density in different points and build an histogram
from where the steady-state PDF of the system can be inferred for different values of the 
control parameters (and correspondingly $\phi_{\tt exp}(x;\sigma)=-\ln\rho_{\tt ss}^{\tt exp}(x;\sigma)$).
\item Second, one follows a given particle during a prescribed protocol (sampling its position)
and build, for the considered
realisation of the experiment, the Hatano-Sasa functional: 
$\mathcal{Y}_{\tt exp}[{\bf x}_{\tt exp};\sigma]=\int_0^\tau dt\dot{\sigma}\frac{\partial\phi_{\tt exp}}
{\partial\sigma}(x_{\tt exp};\sigma)$.
\item Finally, one averages over many experimental realisations the quantity of interest, for example
one determines the average $\langle\exp[-\mathcal{Y}_{\tt exp}[{\bf x}_{\tt exp};\sigma]]\rangle$ in order
to test the validity of the Hatano-Sasa IFT. 
\end{itemize}
Although these settings are very generic, there are  cases where the steady-state PDF of the system
is known, and one do not need to consider this source of experimental errors, as
in \cite{Trepagnier}. 

Let us first note that $\phi_{\tt exp}$ and $x_{\tt exp}$ are not the true values for the quantities
one is trying to determine in the experiment, but are outcomes from measurements. They thus implicitly
carry errors. We can in principle repeat the experiment many times and perform an histogram in order to
calculate the probability of the experimental values of the H-S functional. Imagine the hypothetical
situation in which we are also able to directly obtain the true value of the H-S functional, from where
we can extract for each experiment the corresponding deviation of the measured value of $\mathcal{Y}$ with
respect to the real one: $\mathcal{E}_{\tt exp}[{\bf x}_{\tt exp},x;\sigma]=\mathcal{Y}[{\bf x};\sigma]
-\mathcal{Y}_{\tt exp}[{\bf x}_{\tt exp};\sigma]$. In that case we are able to experimentally build the
joint PDF
\begin{eqnarray}
\label{expPDF}
P(Y_{\tt exp},E_{\tt exp}) &=& \int_{-i\infty}^{+i\infty}\frac{d\lambda_1 d\lambda_2}{(2\pi)^2}
e^{\lambda_1 Y_{\tt exp}+\lambda_2 E_{\tt exp}} G_{\tt exp}(\lambda_1, \lambda_2),\\
\label{GF} 
G_{\tt exp}(\lambda_1, \lambda_2) &=& \overline{\langle e^{-\lambda_1 \mathcal{Y}_{\tt exp}[{\bf x}_{\tt exp};\sigma]
-\lambda_2 \mathcal{E}_{\tt exp}[{\bf x}_{\tt exp},x;\sigma]}}\rangle.
\end{eqnarray}
In equation (\ref{GF}), the brackets denote the thermal average while the overbar denotes the average over
the distribution of experimental measurements errors.
Let the conditional probability of position outcomes be $\mathcal{P}_{\tt err}[{\bf x}_{\tt exp}|x]$,
that is, the probability of obtaining the trajectory of outcomes $x_{\tt exp}(t)$ being $x(t)$ the true trajectory
of the particle. In this case, we can write a mathematical expression for the experimental generating function (GF)
\begin{equation}
\label{average1.1}
G_{\tt exp}(\lambda_1, \lambda_2)=\int\mathcal{D}x\mathcal{D}x_{\tt exp}\mathcal{P}[{\bf x};\sigma]
\mathcal{P}_{\tt err}[{\bf x}_{\tt exp}|x]
\exp\{-\lambda_1 \mathcal{Y}_{\tt exp}[{\bf x}_{\tt exp};\sigma]
-\lambda_2 \mathcal{E}_{\tt exp}[{\bf x}_{\tt exp},x;\sigma]\}.
\end{equation}
Note that, by a suitable rearrangement of terms, we can rewrite the previous expressions as
\begin{equation}\label{average1.3}
  G_{\tt exp}(\lambda_1, \lambda_2)=\int\mathcal{D}x\mathcal{P}[{\bf x};\sigma]
e^{-\lambda_2 \mathcal{Y}[{\bf x};\sigma]}\int\mathcal{D}x_{\tt exp}\mathcal{P}_{\tt err}[{\bf x}_{\tt exp}|x]
e^{-(\lambda_1-\lambda_2) \mathcal{Y}_{\tt exp}[{\bf x}_{\tt exp};\sigma]}.
\end{equation}
Let us introduce the quantity
\begin{equation}
\label{capitalB}
\mathcal{B}[\lambda_1,\lambda_2,x;\sigma]=\int\mathcal{D}x_{\tt exp}\mathcal{P}_{\tt err}[{\bf x}_{\tt exp}|x]
e^{-(\lambda_1-\lambda_2) \mathcal{Y}_{\tt exp}[{\bf x}_{\tt exp};\sigma]}.
\end{equation}
Then equation (\ref{average1.3}) can be written as follows
\begin{equation}
 \label{average1.4}
 G_{\tt exp}(\lambda_1, \lambda_2)=\langle\mathcal{B}[\lambda_1,\lambda_2,x;\sigma]e^{-\lambda_2 \mathcal{Y}[{\bf x};\sigma]}\rangle
=\langle\mathcal{B}[\lambda_1,\lambda_2,x_R;\sigma]e^{-\lambda_2 \mathcal{Y}[{\bf x}_R;\sigma]-\mathcal{S}[{\bf x};\sigma_R]}\rangle_R,
\end{equation}
where the second equality follows from our generalised Crooks-like relation (see part~\ref{ssec:TR}), identifying
the observable $\mathcal{O}[{\bf x};\sigma]$ with the quantity 
$\mathcal{B}[\lambda_1,\lambda_2,x;\sigma]e^{-\lambda_2 \mathcal{Y}[{\bf x};\sigma]}$. 

Now recalling that
$\mathcal{Y}[{\bf x}_R;\sigma]=-\mathcal{Y}[{\bf x};\sigma_R]$, and also that the total trajectory entropy
production splits into the sum of a non-adiabatic contribution (in this case the Hatano-Sasa functional), and an adiabatic
one, and that weighted averages with the exponential of the adiabatic contribution are equivalent to
averages over trajectories given by the dual dynamics, one writes
\begin{equation}
 \label{average1.5}
 G_{\tt exp}(\lambda_1, \lambda_2)=
\langle\mathcal{B}[\lambda_1,\lambda_2,x_R;\sigma]e^{-(1-\lambda_2) \mathcal{Y}[{\bf x};\sigma_R]}\rangle_R^\dagger.
\end{equation}
Let us now analyse the behaviour of the quantity
$\mathcal{B}[\lambda_1,\lambda_2,x;\sigma]$, when a time reversal
operation is performed. For that purpose, let us impose some physically
acceptable and general properties on the conditional error
distribution: we assume that the source of experimental errors
associated to $\mathcal{P}[{\bf x}_{\tt err}|x]$ (the shape of the
distribution) does not depend explicitly on time neither on the form
of the experimental protocol. This error is basically related to some
inherent properties of the measurement apparatus. On the other hand,
we assume that there is no memory in this distribution: the outcome of
the measurement at time $t$ is only related to the true value of $x$
at the same time (this has indeed being implicitly assumed in the
notation for this probability density). We finally assume that this
distribution does not depend on any time derivatives of the involved
variables. From all this requirements, the distribution
$\mathcal{P}[{\bf x}_{\tt err}|x]$ is invariant upon time reversal
\begin{equation}
 \label{average1.6}
\mathcal{P}[{\bf x}_{\tt err}|x]=\mathcal{P}[{\bf x}^R_{\tt err}|x_R].
\end{equation}
We can then write
\begin{equation}
 \label{average1.7}
\mathcal{B}[\lambda_1,\lambda_2,x_R;\sigma]=\int\mathcal{D}x_{\tt exp}\mathcal{P}_{\tt err}[{\bf x}_{\tt exp}|x_R]
e^{-(\lambda_1-\lambda_2) \mathcal{Y}_{\tt exp}[{\bf x}_{\tt exp};\sigma]}.
\end{equation}
Make now the change of variables $x_{\tt exp}=z_R$. Note  that this transformation has a Jacobian
identically equal to one, and that $\mathcal{Y}_{\tt exp}[{\bf x}^R_{\tt exp},\sigma] =
-\mathcal{Y}_{\tt exp}[{\bf x}_{\tt exp},\sigma_R]$,
which in fact, comes from the definition of $\mathcal{Y}_{\tt exp}$. We then have
\begin{eqnarray}
\mathcal{B}[\lambda_1,\lambda_2,x_R;\sigma] &=&
\int\mathcal{D}z_R\mathcal{P}_{\tt err}[z_R|x_R]
e^{-(\lambda_1-\lambda_2) \mathcal{Y}_{\tt exp}[z_R;\sigma]}=\nonumber\\
 &=&\int\mathcal{D}z\mathcal{P}_{\tt err}[z|x]
e^{-\big[(1-\lambda_1)-(1-\lambda_2)\big]\mathcal{Y}_{\tt exp}[z;\sigma_R]},
\end{eqnarray}
which in fact means
\begin{equation}
 \label{average2.1}
\mathcal{B}[\lambda_1,\lambda_2,x_R;\sigma]=
\mathcal{B}[1-\lambda_1,1-\lambda_2,x;\sigma_R].
\end{equation}
Using now (\ref{average1.4}), (\ref{average1.5}), and (\ref{average2.1}), we obtain 
\begin{equation}
 \label{average2.2}
G_{\tt exp}(\lambda_1, \lambda_2)=G_{\tt exp}^{\dagger R}(1-\lambda_1, 1-\lambda_2),
\end{equation}
from where one immediately obtains
\begin{equation}
 \label{average2.3}
P(Y_{\tt exp},E_{\tt exp})/P^{\dagger R}(-Y_{\tt exp},-E_{\tt exp})=e^{Y_{\tt exp}+E_{\tt exp}}.
\end{equation}
From the last expression, or by simply using $\lambda_1=1$, $\lambda_2=0$ in (\ref{average2.2}) we obtain,
for example, the relation
\begin{equation}
 \label{average2.4}
\overline{\langle e^{-\mathcal{Y}_{\tt exp}[{\bf x}_{\tt exp};\sigma]}\rangle}=
\overline{\langle e^{-\mathcal{E}_{\tt exp}[{\bf x},x_{\tt exp};\sigma]}\rangle}_R^{\dagger}.
\end{equation}
From the analysis of $\overline{\langle e^{-\mathcal{E}_{\tt exp}}\rangle}_R^{\dagger}$
for specific cases of experimental errors, one can estimate the dispersion of the experimentally
obtained  $\overline{\langle e^{-\mathcal{Y}_{\tt exp}}\rangle}$ around 
 $\langle e^{-\mathcal{Y}}\rangle=1$. 
\subsubsection{Experimental errors for a driven particle in an harmonic trap}
Let's now consider for concreteness an exactly solvable case
illustrating the validity of the derivation we made above. We focus on
the experiment of Trepagnier \textit{et al}~\cite{Trepagnier}.  In the
considered experimental situation, a microscopic bead is dragged
through water by using a steerable harmonic optical trap. In this
case, treating the bead as a Brownian particle in a harmonic
potential, the steady-state distribution for the system was obtained
theoretically. In the experiment the velocity of the trap plays the
role of the tunable control parameter. For more details on the
experimental setup and on the definitions, see~\cite{Trepagnier}. In
the given case, the Hatano-Sasa functional writes
\begin{equation}
 \label{HStrep}
\mathcal{Y}[{\bf x};v]=\frac{\beta\gamma}{\kappa}\int_0^\tau dt\dot{v}\big(\kappa x+\gamma v\big),
\end{equation}
where $\kappa$ is the trap constant, $\gamma$ is the friction coefficient of the bead in the solution
and $\beta$ is the inverse temperature.

The experimental Hatano-Sasa functional and the error, can then be written as follows
\begin{align}
& \mathcal{Y}_{\tt exp}[{\bf x}_{\tt exp};v]=\frac{\beta\gamma}{\kappa}\int_0^\tau dt\dot{v}\big(\kappa x_{\tt exp}+\gamma v\big),\\
& \mathcal{E}_{\tt exp}[{\bf x},x_{\tt exp};v]=\beta\gamma\int_0^\tau dt\dot{v}\big(x- x_{\tt exp}\big).
\end{align}
We assume that the general conditions we assumed for $\mathcal{P}_{\tt err}[{\bf x}|x_{\tt exp}]
\equiv\mathcal{P}_{\tt err}[{\bf x}-x_{\tt exp}]$ are valid here. Consider now the generating functional of this
conditional probability, which we denote by $G_{\tt err}^J$
\begin{equation}
 \label{trepGJ}
G_{\tt err}^J[J]=\int\mathcal{D}\eta\mathcal{P}_{\tt err}[\eta]e^{-i\int_0^\tau dtJ(t)\eta(t)}.
\end{equation}
Then, rearranging equation (\ref{average1.1}) one has
\begin{eqnarray}
\label{trepG}
\fl  G_{\tt exp}(\lambda_1, \lambda_2)=\int\mathcal{D}x\mathcal{P}[{\bf x};\sigma]
e^{-\lambda_1 \mathcal{Y}[{\bf x};v]}\int\mathcal{D}x_{\tt exp}\mathcal{P}_{\tt err}[{\bf x}-x_{\tt exp}]
e^{-(\lambda_2-\lambda_1) \mathcal{E}_{\tt exp}[{\bf x}-x_{\tt exp};\sigma]}=\nonumber\\
=G_{\tt err}^J\big[-i(\lambda_2-\lambda_1)\beta\gamma\dot{v}\big]\langle e^{-\lambda_1 \mathcal{Y}[{\bf x};v]}\rangle.
\end{eqnarray}
Taking into account that $\dot{v}=-\dot{v}_R$, we have
\begin{equation}
\label{trepG1}
G_{\tt exp}(\lambda_1, \lambda_2)=
G_{\tt err}^J\big[-i\big((1-\lambda_2)-(1-\lambda_1)\big)\beta\gamma\dot{v}_R\big]
\langle e^{-(1-\lambda_1)\mathcal{Y}[{\bf x};v_R]}\rangle^\dagger=G_{\tt exp}^{\dagger R}(1-\lambda_1, 1-\lambda_2), 
\end{equation}
which is the result we have obtained before.

In order to conclude with this section, let us exploit the solvability of this model in order to show that 
indeed there is a link between the apparatus precision and the experimentally relevant measured
quantities. For example, in this case one has
\begin{equation}
 \label{trepG2}
\overline{\langle e^{-\mathcal{Y}_{\tt exp}[{\bf x}_{\tt exp};v]}\rangle}=G_{\tt exp}(1,0)
\equiv G_{\tt err}^J\big[i\beta\gamma\dot{v}\big],
\end{equation}
from where we see that the experimental mean value of the exponential of the
Hatano-Sasa functional, is solely and directly determined by the distribution of measurement errors
in the position of the particle. For example, in the case of an extremely precise measurement, one has
$G_{\tt err}^J=1\Rightarrow\overline{\langle e^{-\mathcal{Y}_{\tt exp}}\rangle}=\langle e^{-\mathcal{Y}}\rangle
\equiv1$. Consider now the case in which the experimental error conditional distribution is Gaussian, \emph{i.e.}
\begin{equation}
 \label{trepG3}
\mathcal{P}_{\tt err}[\eta]\sim \exp\bigg\{-\frac{1}{2\Delta^2}\int_0^\tau dt\eta^2(t)\bigg\}
\Rightarrow G_{\tt err}^J[J]=\exp\bigg\{-\frac{\Delta^2}{2}\int_0^\tau dt J^2(t)\bigg\}
\end{equation}
In this case, one sees that the experimental deviation from the Hatano-Sasa IFT is linked to the accuracy of the measurement apparatus:
\begin{equation}
 \label{trepG4}
-\log\overline{\langle e^{-\mathcal{Y}_{\tt exp}[{\bf x}_{\tt exp};v]}\rangle}=
-\frac{\Delta^2\beta^2\gamma^2}{2}\int_0^\tau dt \dot{v}^2(t)<0,
\end{equation}
which is compatible with the experimental results of~\cite{Trepagnier}.
We learn from the last expression that, at very high temperatures, the ``violation'' of the fluctuation theorem introduced
by the lack of accuracy of the apparatus can be cured by thermal fluctuations. On the other hand, rapidly varying protocols
may induce a remarkable enhancement of the referred ``violation'' factor. 

\subsection{Generalised FDT, identities on correlation functions}
In this section, we use the second equality in (\ref{jointprobs1}) in
order to find some symmetries on correlation functions related to a
generalised fluctuation dissipation theorem (FDT) derived \emph{e.g.}
in
\cite{Richardson,Agarwal,Risken,Falcioni,Parrondo,Maes,LacosteEPL,Lacoste}
(see~\cite{Marconi} for a review).
Let us consider the situation in which the system, initially prepared
in a NESS, is submitted to the variation of its parameters
$\sigma_i(t)=\sigma_i^0+\delta\sigma_i(t)$ in such a way that
$|\frac{\delta\sigma_i(t)}{\sigma_i^0}|\ll1$, with
$\sigma_i^0=\sigma_i(0)$ and $\delta\sigma_i(0)=0$.  In this context,
within the particular form of~\cite{Parrondo}, this generalised FDT
can be written as follows
\begin{equation}
 \label{FDT1}
\bigg\langle\frac{\partial\phi}{\partial\sigma_i}(x(t),\sigma_0)\bigg\rangle=\sum_j
\int_0^t \chi_{ij}(t-t')\delta\sigma_j(t')dt',
\end{equation}
where
\begin{equation}
 \label{FDT2}
 \chi_{ij}(t-t')=\frac{d}{dt}C_{ij}(t-t')=\frac{d}{dt}\bigg\langle\frac{\partial\phi}{\partial\sigma_i}(x(t),\sigma_0)
\frac{\partial\phi}{\partial\sigma_j}(x(t'),\sigma_0)\bigg\rangle_{\tt ss}.
\end{equation}

For a system performing equilibrium dynamics, where detailed balance
holds, this correlation function satisfies the well known Onsager
symmetry relation $C_{ij}(t-t')=C_{ij}(t'-t)\equiv C_{ji}(t-t')$, however, in the most
general case of a NESS, this relation breaks down because of the lack
of detailed balance. However, a similar relation can be easily
obtained for this case: $C_{ij}(t-t')=C^\dagger_{ij}(t'-t)$. The
physical meaning of this expressions is clear.  One is performing a
time reversal operation. When detailed balance holds, there are no
probability currents in the system and this implies invariance upon
time reversal. In this case the system and its dual are the same. When
detailed balance breaks down, there is a finite current in the
steady-state, which is odd under time-reversal. In
this case, the original dynamics is equivalent to  the time reversed dual one
(which as expected has a current opposite to the original one).  This
relation, although conceptually
clear, however, does not seems to be useful: it involves correlations in two different physical systems. The
question which comes now is: does it exist a suitable correlation
function capable to exhibit a symmetry involving only the system under
study? The answer is yes. We now discuss briefly the physical grounds
behind this correlation function.
\subsubsection{Dynamical ensembles and weighted averages}
In a full equivalence with equilibrium thermodynamic ensembles,
one can build trajectory-based ensembles for stochastic non-equilibrium systems, together with quantities equivalent to the
partition function and the free-energy ~--- in a dynamical thermodynamic formalism~\cite{Lecomte}.  For equilibrium systems with
a Hamiltonian $H(C)$ depending on configurations in the phase space, one defines in the canonical
ensemble a free energy $F(\beta)=-(1/\beta)\ln\sum_C\exp\{-\beta H(C)\}$, and averages of observables
can be computed as $\langle y\rangle=\sum_C y(C)\exp\big\{\beta\big(F(\beta)-H(C)\big)\big\}$.
The inverse temperature  $\beta$, is a Lagrange multiplier fixing the mean value of the energy, so that plugging a given
$\beta$ privileges in the sum the configurations compatible with a prescribed value of the mean energy.

Now consider, for a dynamical system, a time extensive functional
$\mathcal{K}[{\bf x}]=\int_0^t dt'f\big(x(t')\big)$. The $s$-ensemble, very
similar to the canonical ensemble in equilibrium statistical
mechanics, is defined by the average
\begin{equation}
 \label{FDT3}
\langle y\rangle^s=\frac{\int\mathcal{D}[{\bf x}]\mathcal{P}[{\bf x}]y[{\bf x}]e^{-s\mathcal{K}[{\bf x}]}}
{\int\mathcal{D}[{\bf x}]\mathcal{P}[{\bf x}]e^{-s\mathcal{K}[{\bf x}]}}.
\end{equation}
This approach has proven to be useful in the study of dynamical phase 
transitions~\cite{GarrahanPRL,GarrahanJPA,Bodineau,Gorissen,Turci}.

The conceptual idea behind the use of the parameter $s$ is the same as in the case of the inverse temperature in canonical
ensembles: in the long time limit, fixing the value of $s$ allows one to calculate averages of functionals,
 not for typical trajectories but
for atypical ones, where the mean value of the time extensive functional $\mathcal{K}[{\bf x}]$ 
is fixed. In other words, weighted averages
select trajectories compatible with a prescribed physical scenario. The presence of this parameter imposes some constraints
on the system in favour of rare events, by introducing a well-chosen bias on the evolution of the system.

Even if the question of dealing experimentally with those biased
trajectories is actually far from being answered in a closed form (see
however~\cite{NemotoSasa}), the procedure is interesting from a
conceptual and numerical~\cite{GKP,LecomteTailleur,GKLT} point of view.

Let us now turn again to our problem. The symmetric nature of the correlation function when the system is in equilibrium
breaks down as detailed balance does for systems with finite currents in the steady-state. However, one can guess that,
if one finds an appropriate time-extensive functional defining weighted averages, the symmetry will be restored
for a given value of the Lagrange multiplier. The only thing we need is to average, in non-equilibrium systems, only over
trajectories compatible with some equilibrium effective dynamics. We deal with this in more detail in what follows.

\subsubsection{Restoring the symmetry of the correlation function using weighted averages}
 
We first combine equation (\ref{jointprobs1}) with equation (\ref{eq:reversingtimeinaverage}) to write
\begin{equation}
 \label{FDT4}
\bigg\langle\mathcal{O}[{\bf x};\sigma]e^{-\lambda_1\Ss_{\na}[{\bf x};\sigma] -\lambda_2\Ss_{\ad}[{\bf x};\sigma]}\bigg\rangle=
\bigg\langle\mathcal{O}[{\bf x}^R;\sigma]e^{-(1-\lambda_1)\Ss_{\na}[{\bf x};\sigma^R] -(1-\lambda_2)\Ss_{\ad}[{\bf x};\sigma^R]}\bigg\rangle^R.
\end{equation}
Let us now choose $\mathcal{O}[{\bf x};\sigma]=\exp\{-\lambda_2\mathcal{F}[{\bf x};\sigma]\}$, with $\mathcal{F}[{\bf x};\sigma]=
\mathcal{F}[{\bf x}^R;\sigma]$. With this we can write
\begin{equation}
 \label{FDT5}
\bigg\langle e^{-\lambda_1\Ss_{\na}[{\bf x};\sigma] -\lambda_2\Ss_{\ad}[{\bf x};\sigma]-\lambda_2\mathcal{F}[{\bf x};\sigma]}\bigg\rangle=
\bigg\langle e^{-(1-\lambda_1)\Ss_{\na}[{\bf x};\sigma^R] -(1-\lambda_2)\Ss_{\ad}[{\bf x};\sigma^R]-\lambda_2\mathcal{F}[{\bf x};\sigma]}
\bigg\rangle^R.
\end{equation}
Let us note now that we can write, to second order in the small perturbations $\delta\sigma$
\begin{eqnarray}
 \label{FDT6}
e^{-\lambda_1\Ss_{\na}[{\bf x};\sigma]}\approx 1-\lambda_1\sum_i\int_0^\tau dt\delta\dot{\sigma}_i(t)
\frac{\partial\phi}{\partial\sigma_i}(x(t),\sigma_0)-\nonumber\\
-\lambda_1\sum_{ij}\int_0^\tau dt\delta\dot{\sigma}_i(t)\delta\sigma_j(t)
\frac{\partial^2\phi}{\partial\sigma_i\partial\sigma_j}(x(t),\sigma_0)+\nonumber\\
 +\frac{\lambda_1^2}{2}\sum_{ij}\int_0^\tau dt dt'\delta\dot{\sigma}_i(t)\delta\dot{\sigma}_j(t')
\frac{\partial\phi}{\partial\sigma_i}(x(t),\sigma_0)\frac{\partial\phi}{\partial\sigma_j}(x(t'),\sigma_0),
\end{eqnarray}
and
\begin{eqnarray}
 \label{FDT7}
e^{-(1-\lambda_1)\Ss_{\na}[{\bf x};\sigma^R]}\approx 1-(1-\lambda_1)\sum_i\int_0^\tau dt\delta\dot{\sigma}^R_i(t)
\frac{\partial\phi}{\partial\sigma_i}(x(t),\sigma_0)-\nonumber\\
-(1-\lambda_1)\sum_{ij}\int_0^\tau dt\delta\dot{\sigma}^R_i(t)\delta\sigma^R_j(t)
\frac{\partial^2\phi}{\partial\sigma_i\partial\sigma_j}(x(t),\sigma_0)+\nonumber\\
+\frac{(1-\lambda_1)^2}{2}\sum_{ij}\int_0^\tau dt dt'\delta\dot{\sigma}^R_i(t)\delta\dot{\sigma}^R_j(t')
\frac{\partial\phi}{\partial\sigma_i}(x(t),\sigma_0)\frac{\partial\phi}{\partial\sigma_j}(x(t'),\sigma_0),
\end{eqnarray}
Plugging  (\ref{FDT6}) and (\ref{FDT7}) in (\ref{FDT5}) and equating the terms with the same power of $\lambda_1$,
we obtain a set of equalities. Let us consider the one corresponding to $\lambda_1^2$:

\begin{eqnarray}
 \label{FDT8}
\fl\bigg\langle\sum_{ij}\int_0^\tau dtdt'\delta\dot{\sigma}_i(t)\delta\dot{\sigma}_j(t')
\frac{\partial\phi}{\partial\sigma_i}(x(t),\sigma_0)\frac{\partial\phi}{\partial\sigma_j}(x(t'),\sigma_0)
e^{-\lambda_2\Ss_{\ad}[{\bf x};\sigma_0]-\lambda_2\mathcal{F}[{\bf x};\sigma_0]}\bigg\rangle_{\tt SS}=\nonumber\\
\fl\bigg\langle\sum_{ij}\int_0^\tau dtdt'\delta\dot{\sigma}^R_i(t)\delta\dot{\sigma}^R_j(t')
\frac{\partial\phi}{\partial\sigma_i}(x(t),\sigma_0)\frac{\partial\phi}{\partial\sigma_j}(x(t'),\sigma_0)
 e^{-(1-\lambda_2)\Ss_{\ad}[{\bf x};\sigma_0]-\lambda_2\mathcal{F}[{\bf x};\sigma_0]}
\bigg\rangle_{\tt SS}.
\end{eqnarray}
In the previous equation, the averages and the functionals in the exponentials can be taken in the steady-state
$\sigma=\sigma_0$, because we are considering only the second order in $\delta\sigma$. In the second term of the previous
equation, we perform the change of variables $t\rightarrow\tau-t$  and $t'\rightarrow\tau-t'$ in the double integral. Then,
the last equation is equivalent to the following identity
\begin{eqnarray}
 \label{FDT9}
\fl\bigg\langle\sum_{ij}\int_0^\tau dtdt'\delta\dot{\sigma}_i(t)\delta\dot{\sigma}_j(t')
\frac{\partial\phi}{\partial\sigma_i}(x(t),\sigma_0)\frac{\partial\phi}{\partial\sigma_j}(x(t'),\sigma_0)
 e^{-\lambda_2\Ss_{\ad}[{\bf x};\sigma_0]-\lambda_2\mathcal{F}[{\bf x};\sigma_0]}\bigg\rangle_{\tt SS}=\nonumber\\
\fl\bigg\langle\sum_{ij}\int_0^\tau dtdt'\delta\dot{\sigma}_i(t)\delta\dot{\sigma}_j(t')
\frac{\partial\phi}{\partial\sigma_i}(x^R(t),\sigma_0)\frac{\partial\phi}{\partial\sigma_j}(x^R(t'),\sigma_0)
 e^{-(1-\lambda_2)\Ss_{\ad}[{\bf x};\sigma_0]-\lambda_2\mathcal{F}[{\bf x};\sigma_0]}
\bigg\rangle_{\tt SS}.
\end{eqnarray}
If we finally take $\lambda_2=1/2$, we obtain a symmetric relation
\begin{eqnarray}
 \label{FDT10}
\fl\bigg\langle\sum_{ij}\int_0^\tau dtdt'\delta\dot{\sigma}_i(t)\delta\dot{\sigma}_j(t')
\frac{\partial\phi}{\partial\sigma_i}(x(t),\sigma_0)\frac{\partial\phi}{\partial\sigma_j}(x(t'),\sigma_0)
 e^{-(1/2)(\Ss_{\ad}[{\bf x};\sigma_0]+\mathcal{F}[{\bf x};\sigma_0])}\bigg\rangle_{\tt SS}=\nonumber\\
\fl\bigg\langle\sum_{ij}\int_0^\tau dtdt'\delta\dot{\sigma}_i(t)\delta\dot{\sigma}_j(t')
\frac{\partial\phi}{\partial\sigma_i}(x^R(t),\sigma_0)\frac{\partial\phi}{\partial\sigma_j}(x^R(t'),\sigma_0)
 e^{-(1/2)(\Ss_{\ad}[{\bf x};\sigma_0]+\mathcal{F}[{\bf x};\sigma_0])}
\bigg\rangle_{\tt SS}.
\end{eqnarray}
From the last expression, we can write
\begin{equation}
 \label{FDT11}
\bigg\langle\mathcal{C}_{ij}(\tau,0)e^{-(1/2)(\Ss_{\ad}[{\bf x};\sigma_0]+\mathcal{F}[{\bf x};\sigma_0])}\bigg\rangle_{\tt SS}
=\bigg\langle\mathcal{C}_{ij}(0,\tau)e^{-(1/2)(\Ss_{\ad}[{\bf x};\sigma_0]+\mathcal{F}[{\bf x};\sigma_0])}\bigg\rangle_{\tt SS},
\end{equation}
where
\begin{equation}
 \label{FDT12}
\mathcal{C}_{ij}(t,t')=\frac{\partial\phi}{\partial\sigma_i}(x(t),\sigma_0)\frac{\partial\phi}{\partial\sigma_j}
(x(t'),\sigma_0).
\end{equation}
Up to now, what we have done is very general and the functional $\mathcal{F}$ can be any quantity being invariant
upon time reversal. In the discussion below, we consider some physically relevant choice of this functional.
\subsubsection{Mapping to an effective system with equilibrium dynamics}
\label{mappingtoeq}
Let us consider the following Langevin evolution
\begin{equation}
 \label{FDT13}
\dot{x} =-\frac{1}{\beta}\frac{\partial \phi}{\partial x}(x;\sigma)+\xi. 
\end{equation}
We know that $\exp\{-\phi\}$ is a well-behaved distribution: it is continuous and
with continuous derivatives at least up to second order. It is also correctly normalised. On the other hand,
this function corresponds to the stationary solution of the Fokker-Planck equation associated
to the process (\ref{FDT13}). This is thus the steady-state of this equation which, if $\phi$ is non
singular in any point of the space (excluding the infinite), is unique. Interestingly, we remark that for the dynamics (\ref{FDT13})
there are no currents in the steady-state, so that detailed balance holds
and the system performs equilibrium dynamics. The steady-state is the same for both
the system (\ref{FDT13}) and the original system under study described by the evolution equation (\ref{langevindynamics}).

The effective action of any process has always a part being odd upon  time reversal,
which is related to entropy production, and an invariant part, which is related to the activity,
which for Langevin dynamics in a potential $\mathcal{V}$ and with additive white noise,
reads
\begin{equation}
 \label{FDT14}
R_{\mathcal{V}}(\tau)=\int_0^\tau dt\bigg[\frac{\beta}{4}\bigg(\frac{\partial\mathcal{V}}{\partial x}\bigg)^2-
\frac{1}{2}\frac{\partial^2\mathcal{V}}{\partial x^2}\bigg].
\end{equation}
The quantity $\exp\big(-\beta dt\frac{dR}{dt}\big)$, is proportional
to the probability that the system has stayed in its configuration
between $t$ and $t+dt$~\cite{Orland}; in other words,
$\beta|\frac{dR}{dt}|$ is the rate at which the system escapes its
configuration, and is related to the activity (or
``traffic'')~\cite{Lecomte,Lefevere,Wynants,PitardEPL}.

Let us investigate how far is our original system from the ``equilibrium'' system (\ref{FDT13}). For that 
we are going to introduce the distance between them in the usual way:
\begin{equation}
 \label{FDT15}
\Xi[{\bf x};\sigma]=\ln\big(\mathcal{P}[{\bf x};\sigma]/\mathcal{P}_\phi[{\bf x};\sigma]\big)=
\frac{1}{2}\bigg(\Ss_{\ad}[{\bf x};\sigma]+ 2\big(R_\phi[{\bf x};\sigma]-R_U[{\bf x};\sigma]\big)\bigg).
\end{equation}
such that $\langle \Xi[{\bf x};\sigma] \rangle \ge 0$ is the Kullback-Leibler distance between $\mathcal{P}[{\bf x};\sigma]$ 
and $\mathcal{P}_\phi[{\bf x};\sigma]$.
This quantity satisfies the corresponding IFT:
\begin{equation}
 \label{FDT16}
\langle e^{-\Xi[{\bf x};\sigma]}\rangle=\langle1\rangle_\phi=1.
\end{equation}
Considering then the $s$-ensemble built using the quantity $\Xi$, we constrain our system to trajectories
compatible with the dynamics (\ref{FDT13}). In particular, averages over the steady-state of the original system
lacking of detailed balance, are directly mapped to an equilibrium steady-state, with zero currents.
Defining this average as
\begin{equation}
\langle\mathcal{O}\rangle^s= \frac{\big\langle\mathcal{O}[{\bf x};\sigma]e^{-s\Xi[{\bf x};\sigma]}\big\rangle}
{\big\langle e^{-s\Xi[{\bf x};\sigma]}\big\rangle},
\end{equation}
we see that averages in the steady-state with $s=1$ are mapped to averages in equilibrium, 
$\langle\ldots\rangle_{\tt ss}^{s=1}\equiv\langle\ldots\rangle_{\tt eq}$. If we take in equation (\ref{FDT11}),
$\mathcal{F}=2(R_\phi-R_U)$, this weighted correlation function acquires
an interesting physical meaning. In order to clarify it a little bit, let us consider the generalised FDT given
by (\ref{FDT1}) in the case where the underlying dynamics is given by (\ref{FDT13}). In order to avoid confusions,
let us clarify the notation even if it will be the same already used. For averages under the original dynamics with
varying protocol $\sigma$, we use the symbol $\langle\ldots\rangle$ whereas for averages under the original dynamics
with constant $\sigma=\sigma_0$, we use the notation $\langle\ldots\rangle_{\tt ss}$. On the other hand, for averages
under the dynamics given by equation (\ref{FDT13}) with varying $\sigma$ we use $\langle\ldots\rangle_\phi$ whereas
for averages under this dynamics for constant $\sigma=\sigma_0$, we use $\langle\ldots\rangle_{\tt eq}$.
We will have
\begin{equation}
 \label{FDT17}
\bigg\langle\frac{\partial\phi}{\partial\sigma_i}(x(t),\sigma_0)\bigg\rangle_\phi=\sum_j
\int_0^t \chi_{ij}^{\tt eq}(t-t')\delta\sigma_j(t')dt',
\end{equation}
where
\begin{equation}
 \label{FDT18}
 \chi_{ij}^{\tt eq}(t-t')=\frac{d}{dt}C_{ij}^{\tt eq}(t-t')=\frac{d}{dt}\bigg\langle\frac{\partial\phi}{\partial\sigma_i}(x(t),\sigma_0)
\frac{\partial\phi}{\partial\sigma_j}(x(t'),\sigma_0)\bigg\rangle_{\tt eq}.
\end{equation}
To the same order in $\delta\sigma$, this implies
\begin{equation}
 \label{FDT19}
\bigg\langle\frac{\partial\phi}{\partial\sigma_i}(x(t),\sigma_0)e^{-\Xi_0(t)}\bigg\rangle=\sum_j
\int_0^t \chi_{ij}^{W}(t-t')\delta\sigma_j(t')dt',
\end{equation}
where
\begin{equation}
 \label{FDT20}
 \chi_{ij}^{W}(t-t')=\frac{d}{dt}C_{ij}^{W}(t-t')=\frac{d}{dt}\bigg\langle\frac{\partial\phi}{\partial\sigma_i}(x(t),\sigma_0)
\frac{\partial\phi}{\partial\sigma_j}(x(t'),\sigma_0)e^{-\Xi_0(t)}\bigg\rangle_{\tt ss},
\end{equation}
and $\Xi_0(t)=\Xi[\mathbf{x};\sigma_0]$.
Let us now introduce the notations $a_i(t)=\frac{\partial\phi}{\partial\sigma_j}(x(t),\sigma_0)$ and 
$b_i(t)=\frac{\partial\phi}{\partial\sigma_j}(x(t),\sigma_0)e^{-\Xi_0(t)}$. Then, one can rewrite the two previous
equations as
\begin{equation}
 \label{FDT21}
\langle b_i(t)\rangle=\sum_j
\int_0^t \chi_{ij}^{ba}(t-t')\delta\sigma_j(t')dt',
\end{equation}
with
\begin{equation}
 \label{FDT22}
 \chi_{ij}^{ba}(t-t')=\frac{d}{dt}C_{ij}^{ba}(t-t')=\frac{d}{dt}\langle b_i(t)a_j(t')\rangle_{\tt ss}.
\end{equation}
Then, equation (\ref{FDT11}) corresponds to the symmetry upon indices interchange:
\begin{equation}
 \label{FDT23}
C_{ij}^{ba}(t-t')=C_{ji}^{ba}(t-t').
\end{equation}

\section{Conclusion}
\label{sec:concl}

We have reviewed, for continuum and discrete systems following a 
Markovian dynamics, how symmetries of the dynamical entropy 
translate into detailed or integrated fluctuation theorems, 
in terms of the symmetries under particular 
transformations of joint probability distributions.

From a conceptual point of view, the FTs given in  (\ref{jointprobs})
provide a convenient perspective to understand the interplay between different
contributions to the total entropy production.  The first of these
equalities shows an explicit relation between the entropy production
fluctuations associated to the system and its environment, valid at
finite time and without quasi-static assumption. It allows one to
study the correlations between the two sources of entropy
production. The second relation unifies the three detailed FTs derived
in~\cite{EspositoPRL}. The relevance of this expression is
twofold. First, it implies the previously known FTs, and secondly, it
gives a relation between the fluctuations of these two different
entropy productions without relying on the dual dynamics (which is
useful from an experimental point of view since one does not need to
run a second system with the dual dynamics). In practice, to compute
the contributions to the total entropy production, one needs to
measure the ratios between transition rates, the PDF of the system at
initial and final times, and the PDF in the steady-state --- which is
achievable (see~\cite{ExperimentsRatios} where ratios between
transitions rates have been measured). In the case of continuous
Langevin equations, only the PDFs have to be measured.  In any case,
 (\ref{jointprobs}) is thus of experimental use.
 
It is known that fluctuation
theorems provide a non-equilibrium way to measure physically relevant
properties of single molecules (such as conformational free energy
differences, see~\cite{ritort} for a review). We have proposed two other outcomes: a way to estimate
experimental errors due to fluctuations in measurement apparatus and a modified non-equilibrium
fluctuation-dissipation relation (FDR)(see
also~\cite{Parrondo,Maes,LacosteEPL,Lacoste}).

In the first case, we have studied the experimental measurements of,
for example, the Hatano-Sasa functional.  We have demonstrated
that the joint PDF of the experimentally measured values and the error
of each measurement satisfies a FT, which allow to extract
precise information about the influence of error sources on the
results.  In particular we have obtained an exact expression for the
``violation factor'' of the Hatano-Sasa IFT in the simple case
considered in the experiment of Trepagnier \textit{et al}~\cite{Trepagnier}. It is worth noting that our
results are compatible with the experimental measurements. We have
explicitly obtained that for protocols with higher dissipation the
experimental results are farther from the expected ones than those of protocols with lower dissipation. This fact is observed in
the broadening of errors bars as dissipation increases. Using the same
ideas, one can study some other family of interesting systems: those
with feedback, where the external protocol imposed to the system
depends on the outcome of some measurement which can be carried out
with some error.

In the second case, we have obtained new symmetries for correlation
functions linked to modified FDRs.  Those symmetries are based on the
use of weighted averages which introduce some bias in the system such
that, controlling this weight, one forces the system to prefer some
trajectories in the phase space. We have indeed shown that, for a
particular form of this weight, the dynamics of the system is exactly
mapped to the dynamics of some equivalent equilibrium system, giving a
physical meaning to these symmetries. Although the result looks
artificial, it can be reformulated in terms of generalised forces and
currents, giving the possibility to recover Onsager reciprocity in the
linear regime in the vicinity of a NESS \cite{us1}.

Last, let us note that joint PDFs-FTs can be relevant in other
contexts.  Consider for example a set of strongly connected systems
which do not satisfy a FT for any of the entropic contributions of its
constituents. In this case, the most precise information one can
obtain involves contributions from all the subsystems.  This can be
seen in classical systems but also in quantum situations. For example,
some recent modified FTs have been obtained for counting-statistics of
electron transport in quantum dots, coupled to quantum point contacts
which play the role of measurements apparatus continuously monitoring
the system (see for example~\cite{QD} for recent developments
and a discussion on the effects of the back action of the environment
on the system). An extension of our procedure to quantum systems may
also give some general framework to tackle such problems \footnote{ For systems
where coherence is not important, described by an effective Pauli
master equation, the mapping to classical systems is
straightforward. Otherwise, we speculate that a Schwinger-Keldysh
approach is relevant.}. 

\ack
This work was supported by CNEA, CONICET (PIP11220090100051), ANPCYT (PICT2007886), 
V.L. was supported in part by the Swiss NSF under MaNEP and Division II and 
thanks CNEA for hospitality. ABK acknowledges Universidad de Barcelona, Ministerio de Ciencia
e Innovaci\'on (Spain) and Generalitat de Catalunya for partial
support through I3 program.

\appendix
\section{Symmetries of operators}
\label{app:symm_operators}

In this appendix we show that the FTs are equivalent for
continuous-time Markov processes to symmetries of the modified
operator of evolution, in the spirit of Lebowitz and
Spohn~\cite{Lebowitz}.

\subsection{Operator approach}

We take the notation of part~\ref{ssec:markov}.
Let's denote $P(\C,\hat h,t)$ the probability of being in
configuration $\C$ at time $t$, having observed a value $\hat h$ of
the reduced Hatano-Sasa functional~\eqref{eq:def_hat_h}, and having started from
configuration $\C_0$. The equation of evolution writes
\begin{equation} \label{eqn:cont_time_Markov_eq_hath}
  \partial_t P(\C,\hat h,t) = 
    \sum_{\C'}\Big[ W(\C'\to\C,\sigma)P\big(\C',\hat h+\phi(\C,\sigma)-\phi(\C',\sigma),t\big)
                                   -  W(\C\to\C',\sigma)P(\C,\hat h,t)  \Big]
\end{equation}
The Laplace transform
$
   \hat P(\C,\lambda ,t) = \int d\hat h\: e^{-\lambda \hat h}P(\C,\hat h,t) 
$
obeys
\begin{equation} \label{eqn:eq_evol_hatP_hs}
  \partial_t \hat P(\C,\lambda ,t) = 
    \sum_{\C'}\Big[ e^{-\lambda [\phi(\C',\sigma)-\phi(\C,\sigma)]}
    W(\C'\to\C,\sigma)\hat P\big(\C',\lambda ,t\big)
                                   -  W(\C\to\C',\sigma)\hat P(\C,\lambda ,t)  \Big]
\end{equation}
or, vectorially $\partial_t|\hat P(\lambda ,t)\rangle = \WW_\lambda  |\hat P(\lambda ,t)\rangle $
with a time-dependent operator $\WW_\lambda $ of elements
\begin{equation}
 \left(\WW_\lambda \right)_{\C,\C'} =
  e^{-\lambda [\phi(\C',\sigma)-\phi(\C,\sigma)]}
    W(\C'\to\C,\sigma) - 
  \delta_{\C,\C'}r(\C,\sigma)
\end{equation}
where $r(\C,\sigma)$ is the escape rate from configuration $\C$, defined as:
\begin{equation}
  r(\C,\sigma)\equiv \sum_{\C'} W(\C\to\C',\sigma)
\end{equation}
Our quantity of interest, $\langle e^{-\lambda \mathcal Y}\rangle\equiv e^{\mu(\lambda ,\tau;\sigma)}$ writes
\begin{align}
 \langle e^{-\lambda  \mathcal Y }\rangle
&= \sum_{\C_0} P_{\st}(\C_0;\sigma_0)
 \int d\hat h\sum_{\C} P(\C,\hat h,\tau) e^{-\lambda \hat h} 
 e^{-\lambda [\phi(\C_0,\sigma_0)-\phi(\C,\sigma_\tau)]}
\\
&= \sum_{\C,\C_0}e^{-\lambda \phi(\C,\sigma_\tau)} \hat P(\C,\lambda ,\tau) e^{-(1-\lambda )\phi(\C_0,\sigma_0)} 
\\
&= \big \langle e^{-\lambda \phi(\C,\sigma_\tau)}
   \big | 
   \mathcal T e^{\int_0^\tau \WW_\lambda } 
   \big|
    e^{-(1-\lambda )\phi(\C,\sigma_0)} \big\rangle 
 \label{eq:ldf-Texp}
\end{align}
where generically $|f(\C)\rangle$ denotes the vector of components $f(\C)$. The time-ordered exponential
\begin{align}
 \mathcal T \exp \left( \int_0^t \mathbb W_\lambda  \right) 
 &= \sum_{n\geq 0} \int_0^t dt_n \int_0^{t_n} dt_{n-1} \ldots
                  \int_0^{t_2} dt_1 \: \mathbb W_\lambda (t_n) \ldots \mathbb W_\lambda (t_1)
\label{eq:Texp_def}
\end{align}
solves $ \partial_t |{\psi(t)}\rangle = \mathbb W_\lambda (t) |{\psi(t)}\rangle$ as easily checked.

\subsection{A primer: the case $\lambda =1$ (original Hatano-Sasa equality)}

From \eqref{eq:ldf-Texp} one has 
$ \langle e^{- \mathcal Y }\rangle =
 \big \langle e^{-\phi(\C,\sigma_\tau)}
   \big | 
   \mathcal T e^{\int_0^\tau \WW_{\lambda =1}} 
   \big|
    1 \big\rangle
$
but
\begin{equation}
 \Big(\WW_{\lambda =1} |1\rangle\Big)_\C = \sum_{\C'}\Big(\WW_{\lambda =1} \Big)_{\C\C'}
= \sum_{\C'} \Big[e^{-[\phi(\C',\sigma)-\phi(\C,\sigma)]}
    W(\C'\to\C,\sigma) - 
  \delta_{\C,\C'}r(\C,\sigma)\Big]=0
\end{equation}
since by definition of steady-state 
\begin{equation}
  \sum_{\C'} \Big[e^{-\phi(\C',\sigma)}
    W(\C'\to\C,\sigma) - 
  \delta_{\C,\C'}r(\C,\sigma)e^{-\phi(\C,\sigma)}\Big]
=0
\end{equation}
which means that $\langle e^{- \mathcal Y }\rangle=1$ as expected.
Note that the derivation we have presented makes no use of
time-discretization or chain rules compared to~\cite{Hatano-Sasa}.

\subsection{Symmetry for the Hatano-Sasa large deviation function}

Starting from \eqref{eq:ldf-Texp} one has (denoting $A^T$ the matrix transpose of $A$)
\begin{align}
  \langle e^{- \lambda  \mathcal Y }\rangle & =
  \big \langle e^{-(1-\lambda )\phi(\C,\sigma_0)} 
   \big | \big(
   \mathcal T e^{\int_0^\tau \WW_\lambda } \big)^T
   \big|
    e^{-\lambda \phi(\C,\sigma_\tau)} \big\rangle 
\label{eq:transposedscalarproduct}
\end{align}
One checks \emph{e.g.} from~\eqref{eq:Texp_def} that the transpose of the time-ordered exponential writes
\begin{equation}
   \big(    \mathcal T e^{\int_0^\tau \WW_\lambda } \big)^T
 = \mathcal T e^{\int_0^\tau (\WW^\R_\lambda )^T}
\label{eq:transposeTexp}
\end{equation}
where $\WW^\R_\lambda $ is the operator of evolution with the time-reversed protocol $\sigma^\R(t)\equiv \sigma(\tau-t)$.
Besides,
\begin{align}
  \big((\WW^\R_\lambda )^T\big)_{\C\C'}
 &=
  \big(\WW^\R_\lambda \big)_{\C'\C}
\\
 &=
  e^{-\lambda [\phi(\C,\sigma^\R)-\phi(\C',\sigma^\R)]}
    W(\C\to\C',\sigma^\R) - 
  \delta_{\C,\C'}r(\C,\sigma)
\\
 &=
  e^{-(1-\lambda )[\phi(\C',\sigma^\R)-\phi(\C,\sigma^\R)]}
    W^\dag(\C'\to\C,\sigma^\R) - 
  \delta_{\C,\C'}r^\dag(\C,\sigma)
\\
\intertext{which shows that}
 (\WW^\R_\lambda )^T  & = (\WW^\R_{1-\lambda })^\dag
\end{align}
Using now~\eqref{eq:transposedscalarproduct} and~\eqref{eq:transposeTexp}, one arrives at
\begin{align}
  \langle e^{-  \lambda  \mathcal Y }\rangle & =
  \big \langle e^{-(1-\lambda )\phi(\C,\sigma^\R_\tau)} 
   \big | 
   \mathcal T e^{\int_0^\tau (\WW^\R_{1-\lambda })^\dag}
   \big|
    e^{-\lambda \phi(\C,\sigma^\R_0)} \big\rangle 
\end{align}
Since the dynamics of rates $W$ and $W^\dag$ have the same steady
state and hence the same $\phi$, we read comparing
with~\eqref{eq:ldf-Texp} that the large deviation function (ldf) at $s
$ for the protocol $\sigma$ is the same as the ldf at $1-\lambda $ for the
time-reversed protocol $\sigma^\R$ and the dual dynamics:
\begin{equation}
  \mu(\lambda ,\tau;\sigma) = \mu^\dag(1-\lambda ,\tau;\sigma^\R)
  \label{eq:sym_HatanoSasa_mu_dual}
\end{equation}

\subsection{Connection with Lebowitz-Spohn-like current}

In~\cite{Lebowitz} Lebowitz and Spohn introduced
a history-dependent observable $Q$ obeying a Gallavotti-Cohen-like symmetry for Markov process with
\emph{time-independent} jump rates. We may generalise their approach by defining, for a given
history and a fixed time-dependent protocol~$\sigma$
\begin{equation}
  Q_\tau = 
      \sum_{k=1}^K \log\frac{W(\C_{k-1}\to\C_k,\sigma_{t_k})}{W^\dag(\C_{k-1}\to\C_k,\sigma_{t_k})}
    = -\hat h_\tau + 
      \sum_{k=1}^K \log\frac{W(\C_{k-1}\to\C_k,\sigma_{t_k})}{W(\C_{k}\to\C_{k-1},\sigma_{t_k})}
\end{equation}
where $\hat h_\tau$ is the reduced Hatano-Sasa defined
in~\eqref{eq:def_hat_h}. Note that when the protocol is
time-independent, the contribution $\hat h_\tau$ to $Q$ merely sums up
to $\phi(\C_K)-\phi(\C_0)$, and one recovers the Lebowitz-Spohn
definition of $Q$. Moreover, detailed balance is equivalent to having
$Q_\tau=0$ for all histories.  One studies here the symmetries of
$\langle e^{-\lambda  \mathcal Y -\kappa Q_\tau}\rangle$.

\bigskip

In the same way as above, one may introduce the probability  $P(\C,\hat h,Q,t)$ of being in
configuration $\C$ at time $t$, having observed a value $\hat h$ of
the reduced Hatano-Sasa functional and a value $Q$ of the Lebowitz-Spohn one ---~having started from
configuration $\C_0$. Following the same steps, one checks that the Laplace transform
\begin{equation} \label{eqn:def_hatP_hs_QLS}
  \hat P(\C,\lambda ,\kappa,t) = \int d\hat h dQ\: e^{-\lambda \hat h-\kappa Q}P(\C,\hat h,Q,t) 
\end{equation}
evolves according to $\partial_t|\hat P(\lambda ,\kappa,t)\rangle = \WW_{\lambda ,\kappa} |\hat P(\lambda ,\kappa,t)\rangle $
with a time-dependent operator $\WW_{\lambda ,\kappa}$ of elements
\begin{equation}
 \left(\WW_{\lambda ,\kappa}\right)_{\C,\C'} =
  e^{-(\lambda -\kappa)[\phi(\C',\sigma)-\phi(\C,\sigma)]}
    W(\C'\to\C,\sigma)^{1-\kappa} W(\C'\to\C,\sigma)^\kappa - 
  \delta_{\C,\C'}r(\C,\sigma)
\end{equation}
One checks that upon transposition the operator possesses the following symmetry:
\begin{equation}
  (\WW_{\lambda ,\kappa})^T   = \WW_{1-\lambda ,1-\kappa}
\end{equation}
Using now that as in~\eqref{eq:ldf-Texp} our quantity of interest
$\langle e^{-\lambda  \mathcal Y -\kappa Q_\tau}\rangle$ writes
\begin{align}
 \langle e^{-\lambda  \mathcal Y -\kappa Q_\tau}\rangle
&= \sum_{\C_0} P_{\st}(\C_0;\sigma_0)
 \int d\hat h dQ \sum_{\C} P(\C,\hat h,Q,\tau) e^{-\lambda \hat h-\kappa Q} 
 e^{-\lambda [\phi(\C_0,\sigma_0)-\phi(\C,\sigma_\tau)]}
\\
% &= \sum_{\C,\C_0}e^{-\lambda \phi(\C,\sigma_\tau)} \hat P(\C,\lambda ,\kappa,\tau) e^{-(1-\lambda )\phi(\C_0,\sigma_0)} 
% \\
&= \big \langle e^{-\lambda \phi(\C,\sigma_\tau)}
   \big | 
   \mathcal T e^{\int_0^\tau \WW_{\lambda ,\kappa}} 
   \big|
    e^{-(1-\lambda )\phi(\C,\sigma_0)} \big\rangle 
 \label{eq:ldf-Texp_withQLS}
\end{align}
and using~\eqref{eq:transposedscalarproduct}, \eqref{eq:transposeTexp}
one obtains that the large deviation function $\mu(\lambda ,\kappa,t)$ defined as
$
  \mu(\lambda ,\kappa,\tau;\sigma)=\log \langle e^{-\lambda  \mathcal Y -\kappa Q_\tau}\rangle
$
possesses the symmetry (at all times)
\begin{equation}
  \mu(\lambda ,\kappa,\tau;\sigma)
=
  \mu(1-\lambda ,1-\kappa,\tau;\sigma^\R)
\label{eq:sym_HatanoSasa_Q_nodual}
\end{equation}

Note first that no reference is made here to the dual dynamics. This also writes
\begin{equation}
  \langle e^{-\lambda  \mathcal Y -\kappa Q_\tau}\rangle =  \langle e^{-(1-\lambda ) \mathcal Y  - (1-\kappa) Q_\tau}\rangle_{\sigma^\R}
\end{equation}
Moreover, focusing on the case $\kappa=0$ one sees that
\begin{equation}
  \langle e^{-\lambda  \mathcal Y }\rangle =  \langle e^{-(1-\lambda ) \mathcal Y  - Q}\rangle_{\sigma^\R}
\end{equation}
which the equivalent of the equality
\begin{equation}
  S_\lambda [{\bf x}^\R;\sigma]=  S_{1-\lambda }[{\bf x};\sigma^\R]+\beta Q_{\hk}[{\bf x};\sigma^\R]
\end{equation}
valid for systems with Langevin dynamics. In the same way, for $\lambda =1$
one sees that
\begin{equation}
  \langle e^{- Q}\rangle= 1 
\end{equation}
This shows that $Q$ plays for Markov chains the same role as $\beta
Q_{\hk}$ for systems with Langevin dynamics.

\break

\end{document}